\documentclass{amsart}

\usepackage{fancyhdr}
\usepackage{amsmath} 
\usepackage{amsthm} 
\usepackage{cite}
\usepackage{hyperref} 
\usepackage{graphics} 
\usepackage{algorithmic} 
\usepackage{algorithm} 
\usepackage{tikz}
\usepackage{subfig}
\usepackage{url}

\usetikzlibrary{shapes,arrows}

\newenvironment{nouppercase}{%
	\renewcommand{\uppercasenonmath}[1]{}}{}

\begin{document}

\title[Center-Based Attacks]{Attack Vulnerability of Complex Networks in Center-Based Strategies}


\author[Lekha et.al.]{
{\sc Divya Sindhu Lekha}$^*$,\\[2pt]
Department of Computer Applications, Cochin University of Science and Technology,Kochi, India \\
$^*${Corresponding author. Email: divi.lekha@gmail.com}\\[2pt]
{\sc Kannan Balakrishnan}\\[2pt]
Department of Computer Applications, Cochin University of Science and Technology,Kochi, India\\
{Email: mullayilkannan@gmail.com}\\[6pt]
}

\begin{nouppercase}
\maketitle
\end{nouppercase}

\begin{abstract}
{Central nodes are critical in establishing structural connectivity in a complex network. Attacking such nodes can create real havoc in a complex system. We propose attack strategies based on four types of centers, namely betweenness center, degree center, median, and center. We study the vulnerability of synthetic as well as real-world networks in these \emph{center-based attacks}. These attacks are node-removal attacks which involve identifying the central node set and removing them from the network. We observed that the attacks based on recalculated network information are more efficient than ones based on initial network information. This work shows that the median-based attack, which is a novel strategy proposed in this work, is highly destructive in real as well as synthetic networks.}
{Complex networks, Vulnerability, Graph theory, Center-based attacks, Recalculated attacks.}
\\
2000 Math Subject Classification: 34K30, 35K57, 35Q80, 92D25
\end{abstract}

\section{Introduction}
Analysis and prediction of behavioural dynamics of complex systems require a thorough study of their underlying network structures. Real-life networks like biological networks handle a massive amount of data. Such big-data networks are becoming larger and complex over time. And so are the tools for modelling and analyzing them. Network science is one of those scientific disciplines which has been emerged by converging the commonalities in many multi-disciplinary research fields. It is a highly specialized as well as an inter-disciplinary subject.  One way for studying real networks is by modelling approximate structures. Graph theory enables such a modelling. The well-studied theories and methods in graph theory can aid in the mathematical description of large and complex networks. Network analysis has found new directions with the incorporation of graph-theoretic concepts in such studies and the advent of complex network theory.

Network security is an all-time relevant research topic. Security threats on complex networks can be either random or targeted. An attacker can either remove or default a functioning node or edge in the network. In targeted attacks, the attacker needs to identify the potential nodes/edges in the network and destroy them. Based on the type of entity being attacked we can term the attacks as node-targeted or edge-targeted attacks. Analyzing attack vulnerability in synthetic networks (network models) can give insights into how real-world networks behave in such attacks. Such studies are highly recommendable while designing robust and error-tolerant networks. While studying vulnerability, it is always desirable to compare the behaviour of synthetic and real-world networks. Here arise the need for a uniform metric for network vulnerability. Size of the largest connected component ($LCC$) in the network can be considered as one such metric. 
In this paper, we focus on the vulnerability of synthetic as well as real-world networks in \emph{center-based attacks}. Center-based attacks are node-removal attacks which involve identifying the central node set in the initial network and removing them from the network. Four types of centers are considered here. They are 
\begin{enumerate}
	\item Betweenness center $BC$
	\item Center $C$ 
	\item Degree center $DC$ and
	\item Median $M$
\end{enumerate}
First, we present a roadmap of the related works in vulnerability analysis of complex networks (Section~\ref{sec: sci}). Familiarisation of the terminologies associated with this study (Section~\ref{sec:term}) follows. The attack strategy is defined in Section~\ref{sec:Strategy}. In the following section (Section~\ref{sec: Synth}) we analyse the robustness of synthetic networks. The network models considered here are:
\begin{enumerate}
	\item Random Network - Erd\H{o}s-R\'{e}nyi (ER) model
	\item Small-world Network - Watts-Strogatz model
	\item Scale-free Network - Barab\'{a}si-Albert model
\end{enumerate}
In the next section (Section~\ref{sec: Real}) we analyse the vulnerability of real-world networks. The networks are chosen from different domains areas. They are
\begin{enumerate}
	\item Collaboration Network - Network Science co-authorship network~\cite{New06}
	\item Literature Network - Les Miserables co-appearance network~\cite{Knu93}
	\item Airport Network - US popular airports connection network~\cite{CPV07}
	\item Biological Network - Protein interaction network in Yeast~\cite{JMB01}
\end{enumerate}
\section{Related Works}\label{sec: sci}
Research in vulnerability analysis of complex networks has been particularly active during the last decade.  Albert et al. in 2000~\cite{AJB00} studied the error and attack tolerance of complex networks. In this work, they observe that scale-free networks show an unexceptionally high level of robustness. At almost the same time, Holme et al. (2002)~\cite{HKY02} observed that the behaviour of real-world networks towards node-attacks was very much different from that of network models. Therefore they suggest that there might be other structural phenomena which contribute to the behaviour of real networks in attacks.
Motter and Lai (2002)~\cite{ML02} studied the security threat to complex networks by the cascade of failures due to intentional attacks. Their findings underline that many real-world networks which are naturally heterogeneous (and survive random attacks) are in fact highly prone to such cascading failures. Later Motter himself proposed a strategy for defending such attacks in heterogeneous networks~\cite{Mot04} in 2004. Following these findings, Crucitti et al. in their work~\cite{CLM04} introduced a model which depicts that attack on a single node is sufficient to destroy the efficiency of the entire network. 

Attack vulnerability is also an indication of many structural and dynamic features of networks. Goldshtein et al.~\cite{GKS04}, in 2004, studied the hierarchy of complex networks based on its vulnerability measures. Thus the quantification of vulnerability measures turns out to be of relevance in network analysis. Boccaletti et al., in 2007, proposed measures to quantify the multiscale evaluation of vulnerability~\cite{BBC07}. Mishkovski et. al.~\cite{MBK11}, in 2011, proposed a metric for vulnerability measurement. This metric was based on average edge betweenness. 

Investigations on the behaviour of complex networks towards different types of attacks flourished recently. One notable contribution among them is the node-removal strategy proposed by Bellingeri et al. (2014)~\cite{BCV14}. They observed that the efficiency of strategies depends very much on the network topology. Another closely related contribution is by Nie et al. (2015)~\cite{NGZ15} in which they proposed two new attack strategies based on both degree and betweenness. They found that these strategies were highly efficient than traditional ones based on degree alone. Attack strategies based on paths were introduced recently. Pu and Cui (2015)~\cite{PW15} proposed attack strategies based on longest paths and observed that homogenous networks are fragile in this category of attacks. Another most recent work in this regard is the shortest path-attacks proposed by Hao et al. (2016)~\cite{HHC16}. They found that small world models are highly robust towards such attacks.

During each iteration of an attack, the network attributes undergo tremendous changes. The initial network attributes may no more remain invalid. Therefore after each iteration of attack, it is desirable to recalculate the network measures based on which the attack is proposed. In 2002, Holme et al. (2002)~\cite{HKY02} observed that the attack strategies based on recalculated information are more harmful than the strategies based on initial information. While investigating the identification of vital nodes in complex networks, Lu et. al.~\cite{LCR16} also noticed that the strategies based on adaptive recalculation are more efficient than the straightforward methods.

Real-world networks are mostly weighted networks. Vulnerability analysis of weighted networks is a new area of research. Bellingeri and Cassi, in their new paper~\cite{BC18} (2017), focus on the change in efficiency of attacks due to weights in the networks. Their work suggests that the vulnerability analysis without considering the weighted structure of networks could be at times misleading.

\section{Terminology}\label{sec:term}
\subsection{Graph distances and centers}
Two vertices $u$ and $v$ in a graph $G$ are said to be adjacent if there is an edge $(u, v)$ joining $u$ and $v$. The order of $G$ is defined as $n = |V|$ and size of $G$ as $m = |E|$. 
\emph{Distance} between two vertices $u$ and $v$ in a graph, denoted by $d(u, v)$ is the length of a $u-v$ geodesic (shortest path between $u$ and $v$).
\begin{description}
	\item[Eccentricity and center]  Eccentricity of a vertex $v$ on $G$ is defined as $$E_G(v) = \max_{u \in V}d_G(u, v)$$ where $d_G(u, v)$ is the length of a shortest $u-v$ path in G. A center vertex is a vertex with minimum $E_G(v)$. The set of all center vertices is the center set of $G$ and is denoted by $C(G)$~\cite{BH90}.
\end{description}
\begin{description}
	\item[Betweenness and betweenness center] Betweenness of vertes $v$ in $G$ denoted by $B_G(v)$ or simply $B(v)$  may be defined as $$B(v) = \sum_{s,t \in V(G)\setminus\{v\}}{\frac{\sigma_{st}(v)}{\sigma_{st}}}$$ where  $\sigma_{st}(v)$ denotes the number of shortest $s$-$t$ paths  passing through $v$ and $\sigma_{st}$, the number of shortest $s$-$t$ paths. The betweenness center of a graph $G$ is a vertex with minimum $B_G(v)$. The set of all betweenness center vertices is the betweenness center set of $G$ and is denoted by $BC(G)$~\cite{Fre78}.
\end{description}
\begin{description}
	\item[Remoteness and median] Remoteness of vertex $v$ on $G$ is defined as $D_G(v) = \sum_{u \in V}d_G(u, v)$ where $d_G(u, v)$ is the length of a shortest $u-v$ path in G. A median vertex is a vertex with minimum $D_G(v)$. The set of all median vertices is the median set of $G$ and is denoted by $M(G)$~\cite{BH90}.
\end{description}
\begin{description}
	\item[Degree and degree center] Degree of vertex $v$ on $G$, $\delta(v)$, is the number of edges incident to the vertex $v$. Degree center of a graph $G$ is a vertex with maximum $\delta(v)$. The set of all degree center vertices is the degree center set of $G$ and is denoted by $DC(G)$.
\end{description}
Graph center and median represent global centers of the graph; whereas a degree center can be considered as a local center. Another center considered in this study is the betweenness center, which is related to the geodesics in the graph. To get an insight into the relevance of graph distances in complex networks see~\cite{GO10} and refer~\cite{WuS03} for more details on center notions in complex networks.
\subsection{Network models}
Theoretic network models studied by mathematicians are mostly random models. But in reality, networks behave in a crucially different manner than random networks. Small-world and scale-free networks are two fundamental models capturing specific architectural features of real networks.
\subsubsection{Random networks}
The classical model is the Erd\H{o}s-R\'{e}nyi model (E-R model) proposed in~\cite{ErR59}. Degree distribution in E-R graphs is binomial which becomes Poisson when $n$ is very large ($n \rightarrow \infty$). The probability of occurance of a vertex with degree $k$ in E-R network is given by $P(k)\approx e^{-k}\frac{\delta^k_{av}}{k!}$
\subsubsection{Small world networks}
Small world problem was first investigated by Milgram in his pioneer work~\cite{Mil67}. He observed that an acquaintance chain between any two people is always embedded in a small-world structure (maximum six people). Most complex networks exhibit smaller diameter. The diameter $dia$ increases logarithmically with the size of the network. i.e. $dia \approx log n$ as $n \rightarrow \infty$~\cite{Erc15}. The most acclaimed model in this regard is the W-S model proposed by Watts and Strogatz in~\cite{WaS98}.
\subsubsection{Scale free networks} 
This model characterizes the dynamicity of an evolving network. Barab\'{a}si and Albert~\cite{AlB99} proposed B-A model with \emph{scale-free property}. This property states that degree distribution is characterized by power-law: $P(k) \propto k^{-\gamma} \textnormal{ with } \gamma > 1$. Evolution of such a model is governed by the \emph{law of preferential attachment}. By this law, the newly added nodes prefer to connect with nodes of higher degree (\emph{hubs}). Thus the degree of hub nodes continuously increases with the addition of new nodes.
\section{Center-based attack: Strategy}\label{sec:Strategy} 
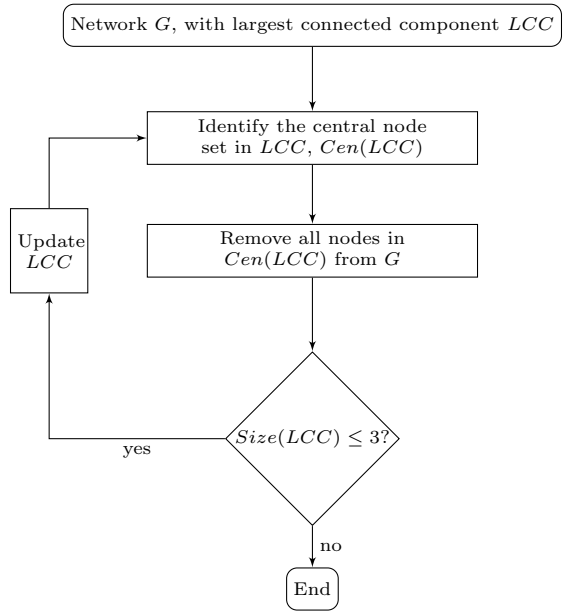
\begin{figure}[h]
	\centering
	\scriptsize
	
	\tikzstyle{decision} = [diamond, draw,  text badly centered, inner sep=0pt]
	\tikzstyle{block} = [rectangle, draw, text width=15em, text centered, minimum height=2em]
	\tikzstyle{block1} = [rectangle, draw, text width=3em, text centered, minimum height=4em]
	\tikzstyle{line} = [draw, -latex']
	\tikzstyle{cloud} = [draw, rectangle, node distance=1.5cm, rounded corners,minimum height=2em]
	
	\begin{tikzpicture}[node distance = 1.5cm, auto]
	\node [cloud] (expert) {Network $G$, with largest connected component $LCC$};
	\node [block, below of=expert,node distance=1.5cm] (identify) {Identify the central node set in $LCC$, $Cen(LCC)$};
	\node [block, below of=identify,node distance=1.5cm] (evaluate) {Remove all nodes in $Cen(LCC)$ from $G$};
	\node [block1, left of=evaluate, node distance=3.5cm] (update) {Update $LCC$};
	\node [decision, below of=evaluate, node distance =2.5cm] (decide) { $Size(LCC) \le 3$?};
	\node [cloud, below of=decide, node distance=2cm] (system) {End};
	\path [line] (expert) -- (identify);
	\path [line] (identify) -- (evaluate);
	\path [line] (evaluate) -- (decide);
	\path [line] (decide) -| node [near start] {yes} (update);
	\path [line] (update) |- (identify);
	\path [line] (decide) -- node {no}(system);
	\end{tikzpicture}
	\caption{The flow diagram of center-based attack}
	\label{fig:flow}
\end{figure}

We propose center-based attacks on complex networks in which targeted entities are the different centers as defined in Section~\ref{sec:term}. Given a network $N$, first step of the strategy is to identify the node sets in the descending level of their centrality ($BC$/ $C$/ $DC$/ $M$) in $N$. An attack is performed by removing all nodes in each identified set and the edges incident on them from $N$. The damage caused to $N$ in each attempt of node removal is calculated in terms of the size of the largest connected component $LCC$ in the updated $N$. In the next iteration, the node set with next higher centrality is attacked. In order to enable the comparison of different networks, we use a normalized value $LCC'$ ~\cite{HHC16} defined as $\frac{LCC}{N}$ where $N$ is the size of the initial network. The attack iterations terminate when the network is completely destructed (when $LCC \le 3$). The proposed attack strategies are:
\begin{enumerate}
	\item Initial betweenness center attack $IB$
	\item Initial center attack $IC$ 
	\item Initial degree center attack $ID$ and
	\item Initial median attack $IM$
\end{enumerate}
These strategies are purely based on the information of the initial network. To achieve more efficient results, it is desirable to follow an approach based on the recalculated information. In such an approach, the central node set is re-identified in the $LCC$ after each attack iteration and removed in the next iteration.  The attacks based on recalculated centrality measures are named as follows:
\begin{enumerate}
	\item Recalculated betweenness center attack $RB$
	\item Recalculated center attack $RC$ 
	\item Recalculated degree center attack $RD$ and
	\item Recalculated median attack $RM$
\end{enumerate}
The strategy is depicted in Algorithm~\ref{algo:cattack} and the following flowchart (figure~\ref{fig:flow}).
\bigskip
\begin{algorithm}[h]  
	\begin{algorithmic}[1]  
		\REQUIRE Network $G$ with $n$ nodes and largest connected component $LCC$
		\ENSURE Disconnected Network
		\STATE Identify the central node set in $L$, $Cen(LCC)$
		\STATE Remove all nodes in $Cen(LCC)$ from $G$.
		\STATE Find the current largest connected component and update $LCC$
		\STATE Determine if $Size(LCC) \le 3$. If true, terminate; else go to step $1$
	\end{algorithmic}  
	\caption{Center-based attack strategy}  
	\label{algo:cattack}  
\end{algorithm}   
\bigskip
\normalfont
\section{Vulnerability analysis of synthetic networks}\label{sec: Synth}
Now we analyse the vulnerability of different network models towards center-based attacks. Many algorithms exist for generating network models (Refer~\cite{AlB02, BB05, DoM02, New00, Str01} for details). We used Pajek software~\cite{BM03, BM98} (\emph{Batagelj, V., Mrvar, A.}: \href{http://vlado.fmf.uni-lj.si/pub/networks/pajek/}{Pajek - Program for Large Network Analysis.}) for generating different network models. Random networks were generated using E-R model with Poisson distribution, small world networks using W-S model and scale-free networks using B-A  model. The size of networks considered were $500$ with average vertex degree of $6$. When the networks generated were disconnected; we extracted the giant component alone from the network for our experiments. Therefore the ultimate aim of the attacks was the complete destruction of the initial giant component of the network.

\begin{figure}[!t]
	\centering
	\subfloat[IM Attack]{\includegraphics[width=2in]{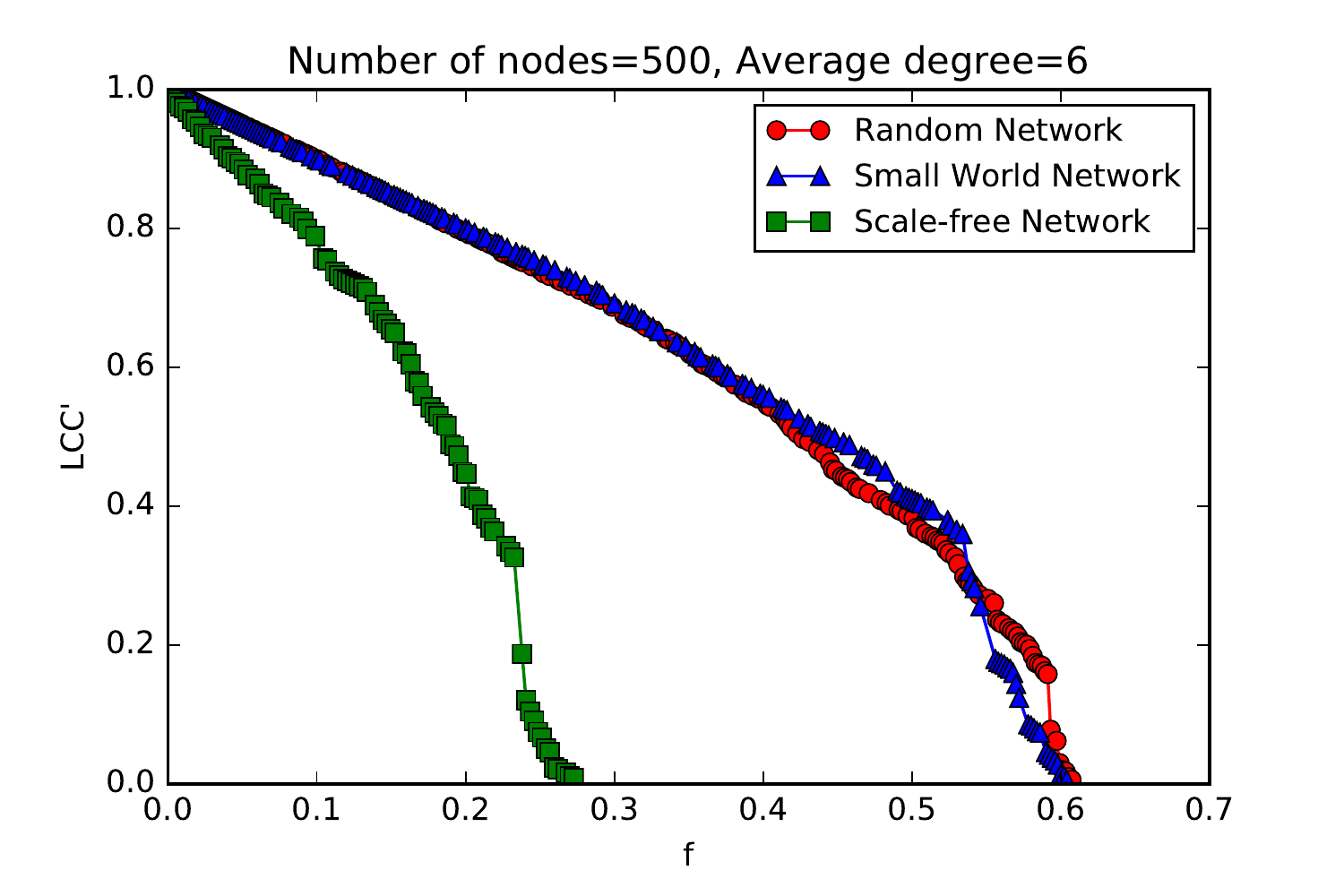}
		\label{fig: IM}}
	\hfil
	\subfloat[IC Attack]{\includegraphics[width=2in]{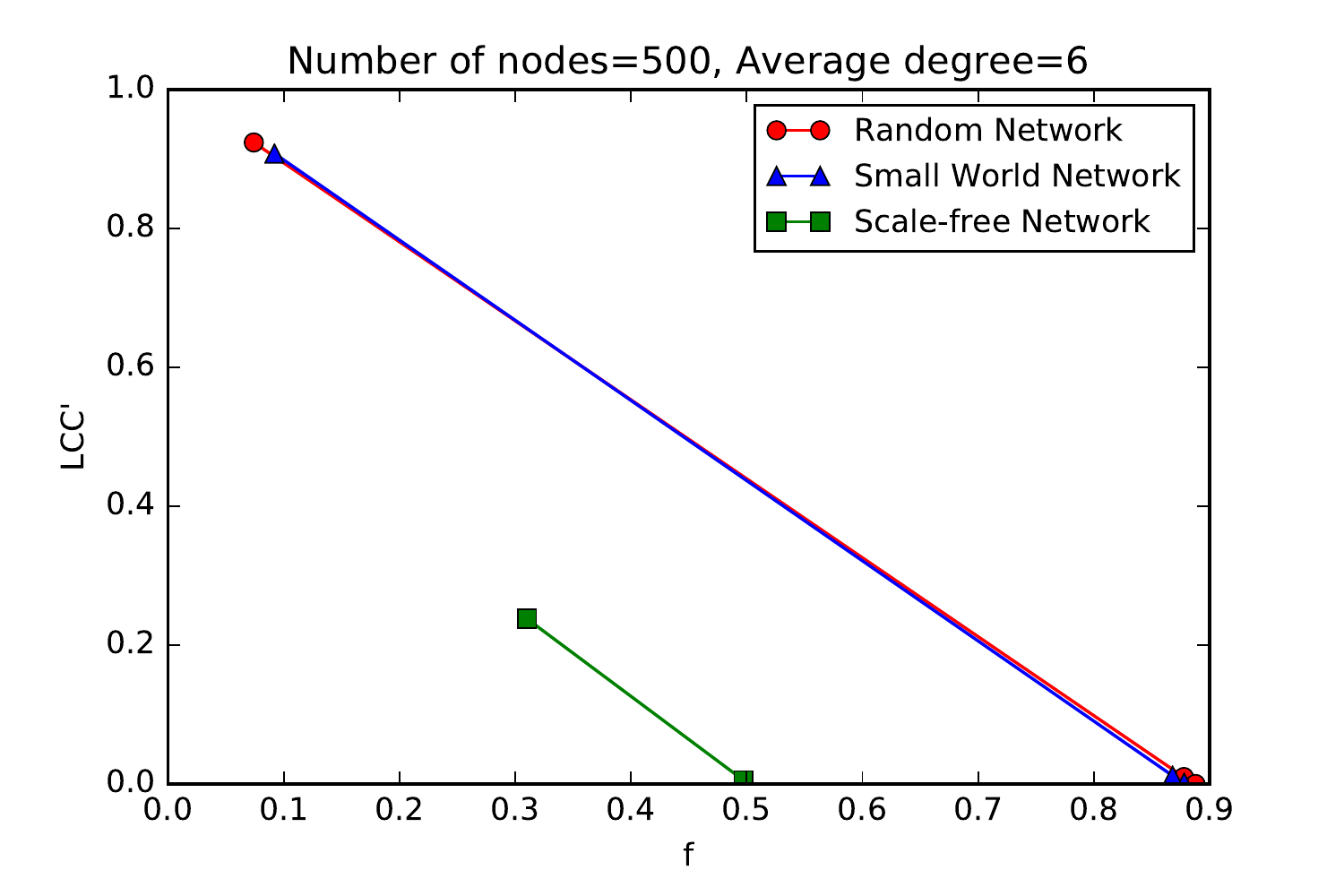}
		\label{fig: IC}}
\\
	\subfloat[IB Attack]{\includegraphics[width=2in]{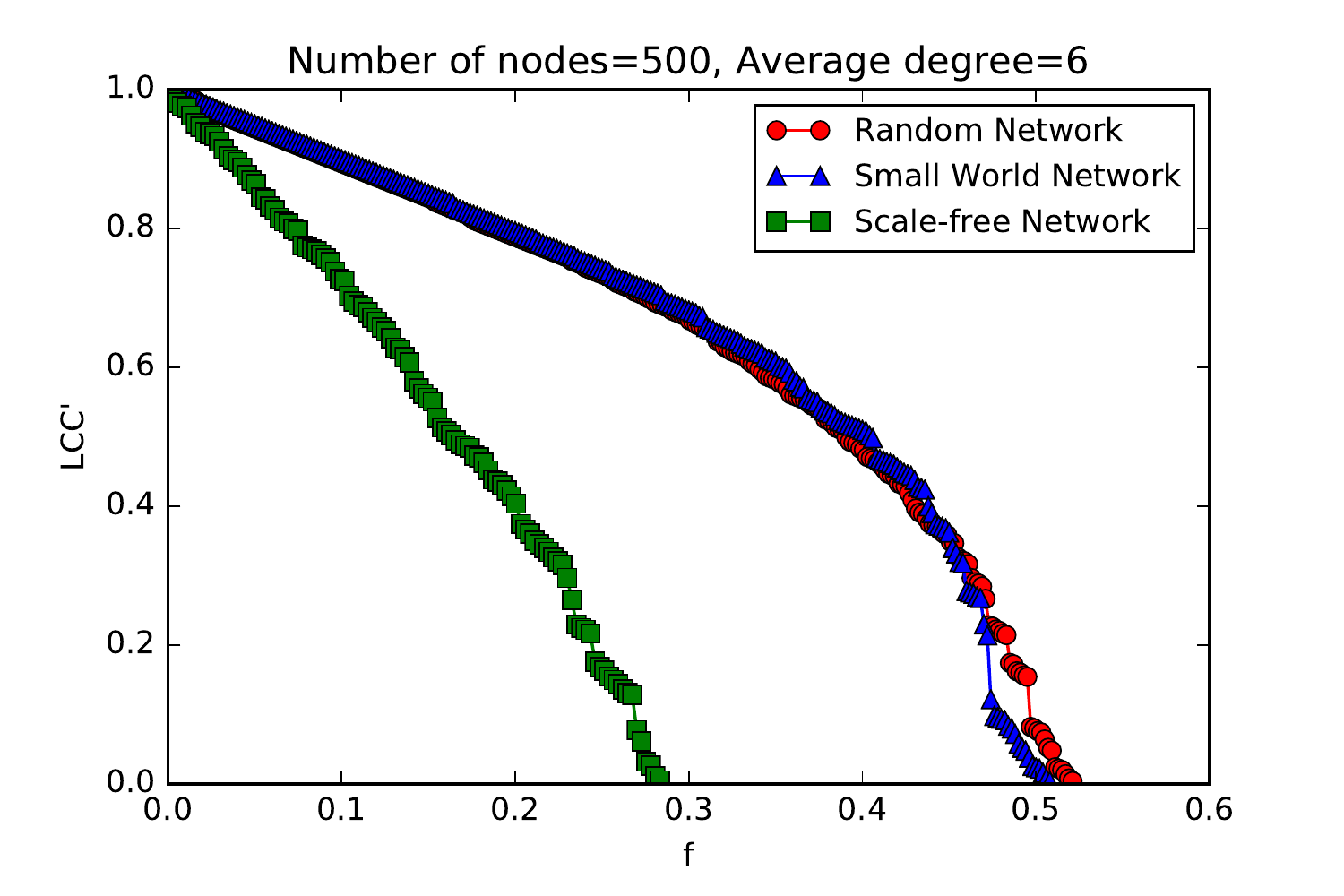}
		\label{fig: IB}}
	\hfil
	\subfloat[ID Attack]{\includegraphics[width=2in]{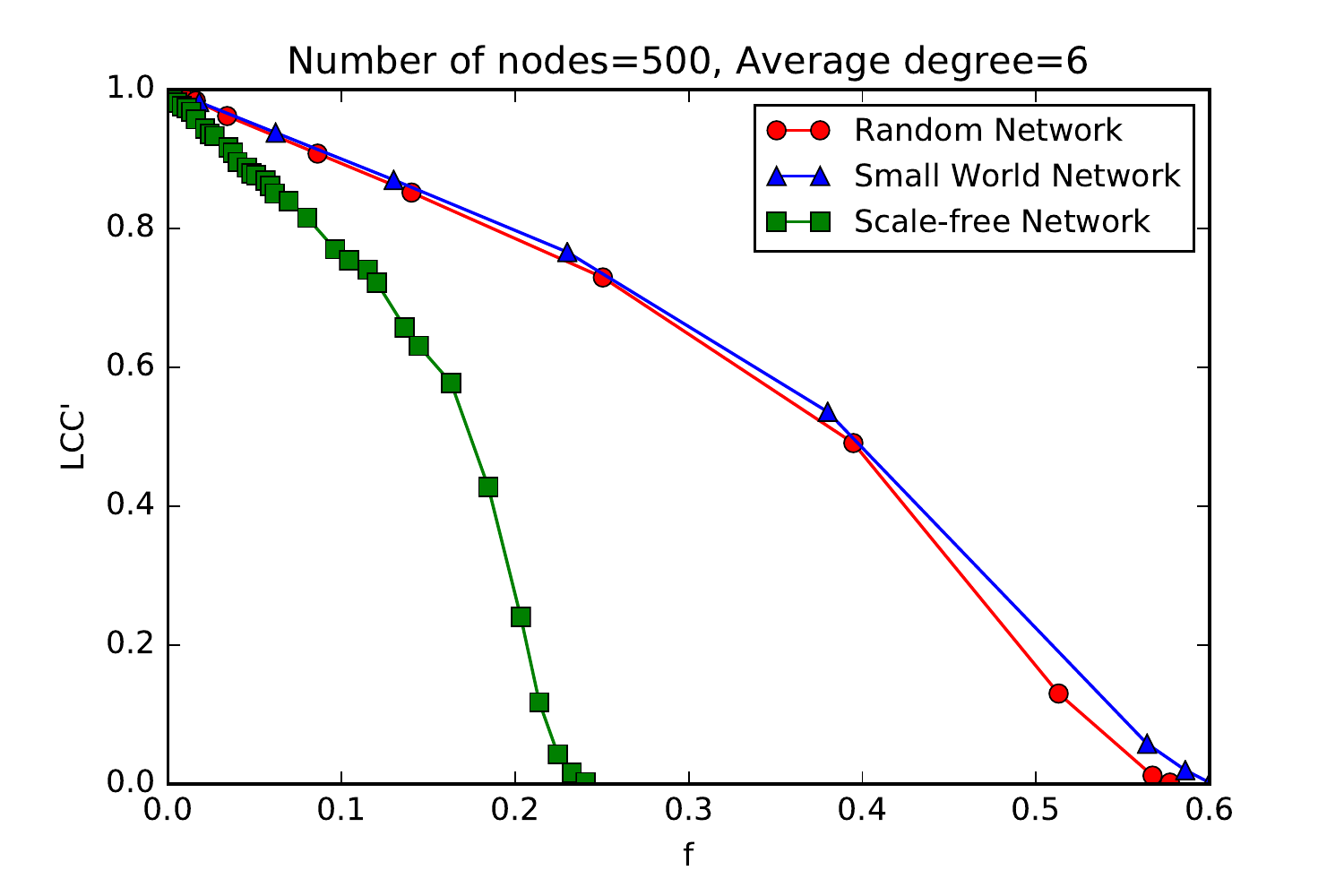}
		\label{fig: ID}}
	\caption{Initial center attacks in different network models.}
	\label{fig_init}
\end{figure}

\begin{figure}[!t]
	\centering
	\subfloat[Random Network]{\includegraphics[width=2in]{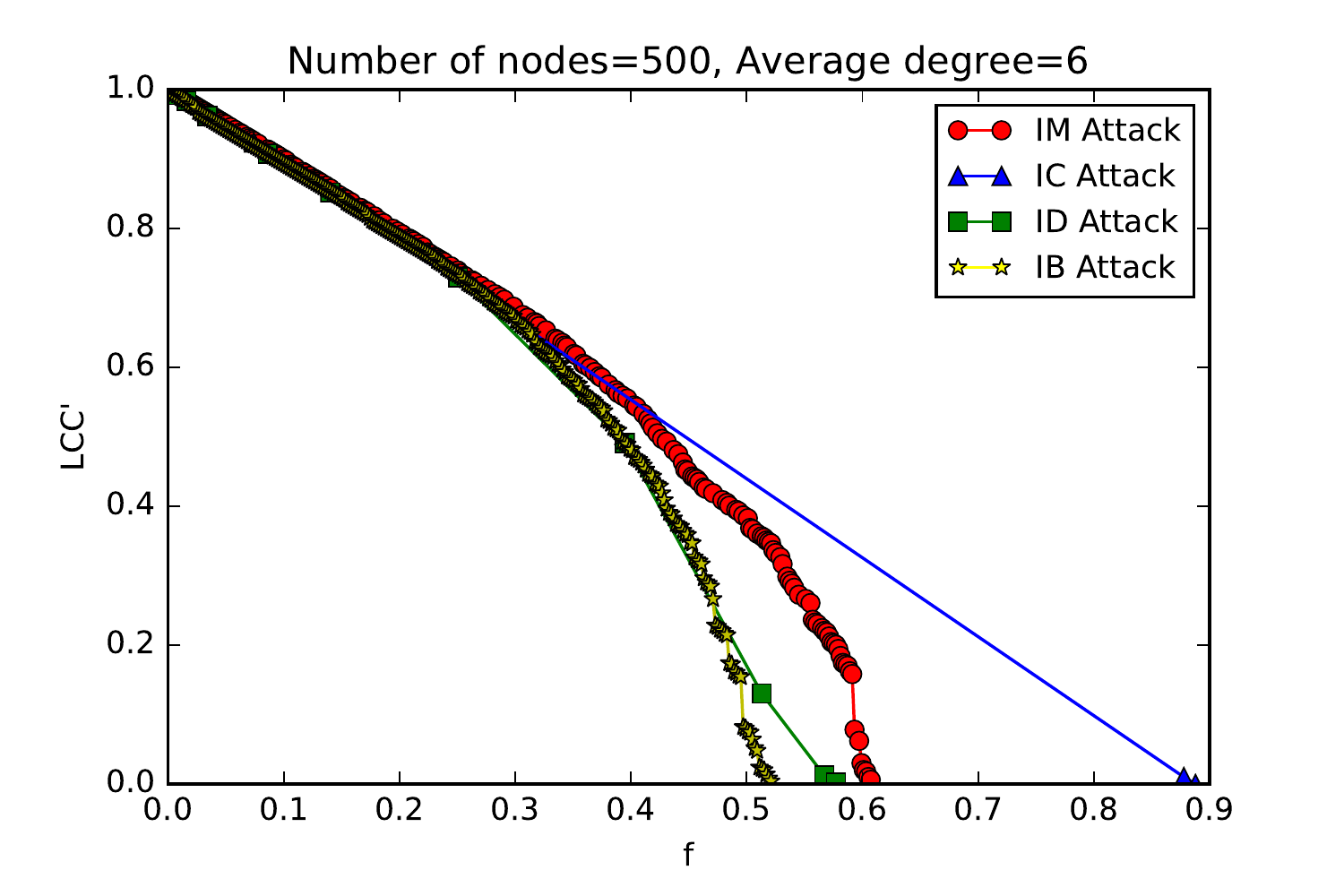}
		\label{fig: R}}
	\hfil
	\subfloat[Scale-Free Network]{\includegraphics[width=2in]{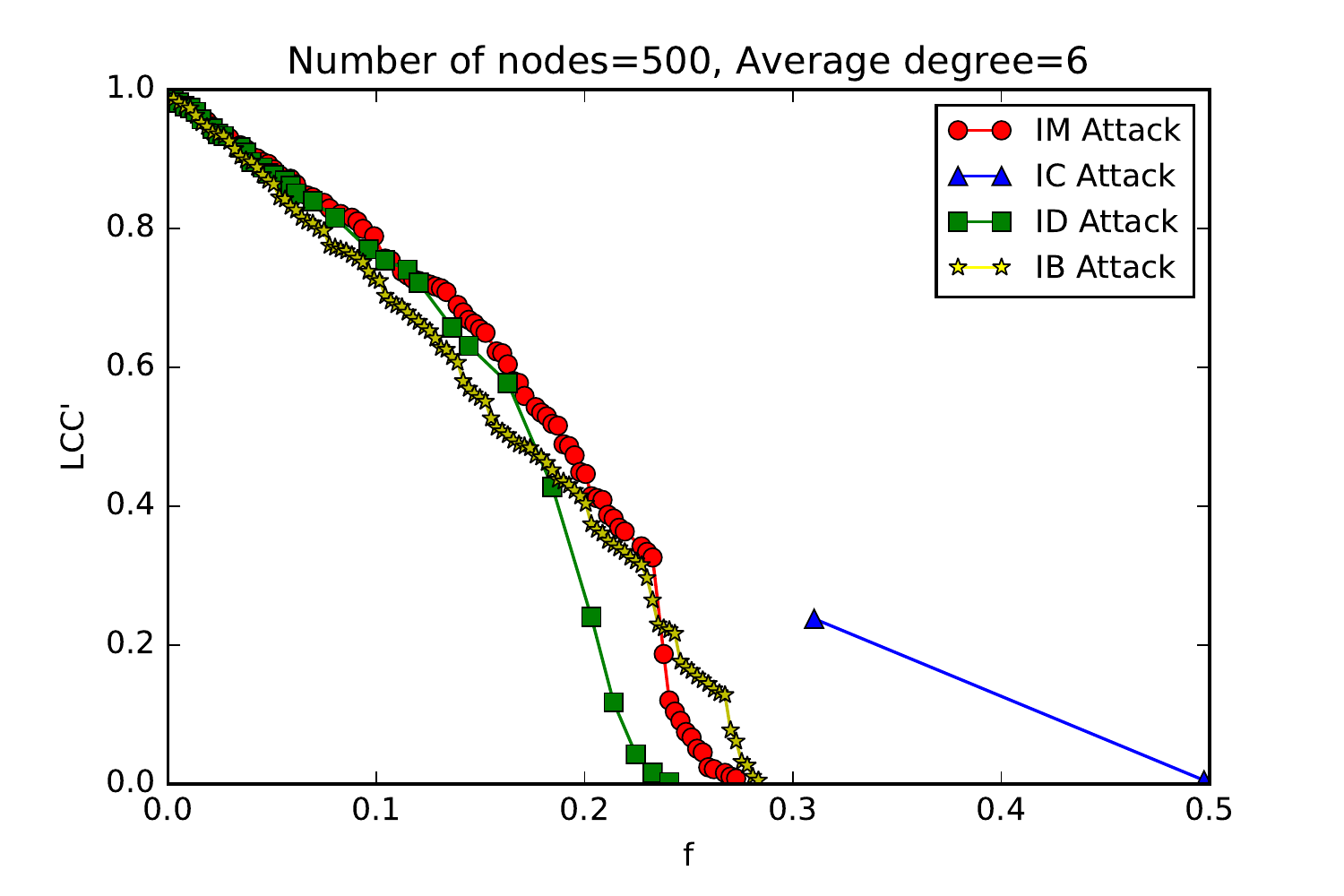}
		\label{fig: SFi}}
\\
	\subfloat[Small World Network]{\includegraphics[width=2in]{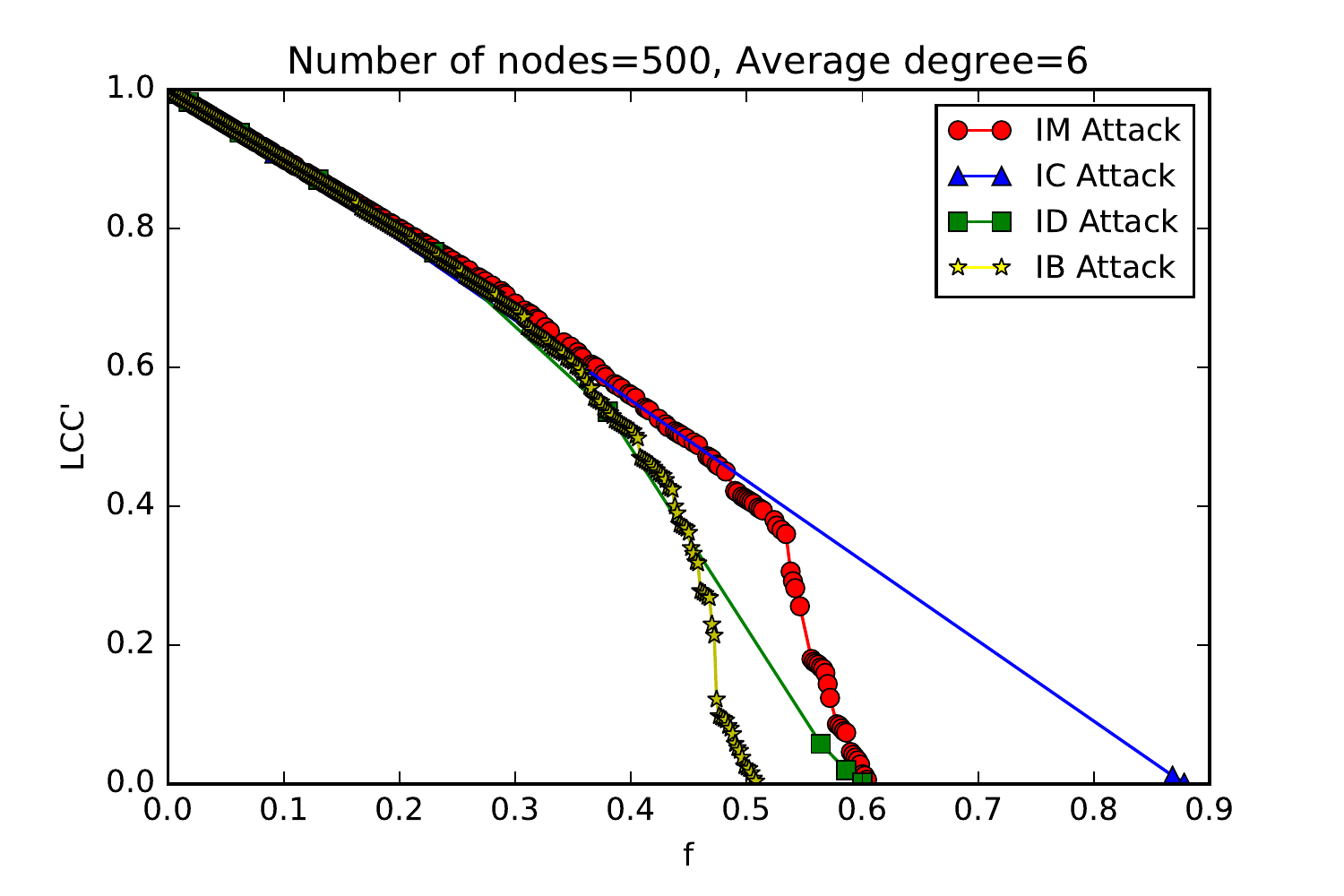}
		\label{fig: SWi}}
	\caption{Vulnerability of network models to initial center attacks.}
	\label{fig_initModel}
\end{figure}

After constructing the networks using generative models, trials of attacks based on the initial network information ($IM, IC, IB, ID$) were carried out. The change in $LCC'$ was analysed against the fraction of nodes removed in the attacks $f$ and compared for all the network models described above. See figure~\ref{fig_init} for a comparison of the strategies $IM, IC, IB$ and $ID$ in the synthetic networks.
The vulnerability of the three network models to different attacks are compared and depicted in figure~\ref{fig_initModel}.

\begin{figure}[!t]
	\centering
	\subfloat[RM Attack]{\includegraphics[width=2in]{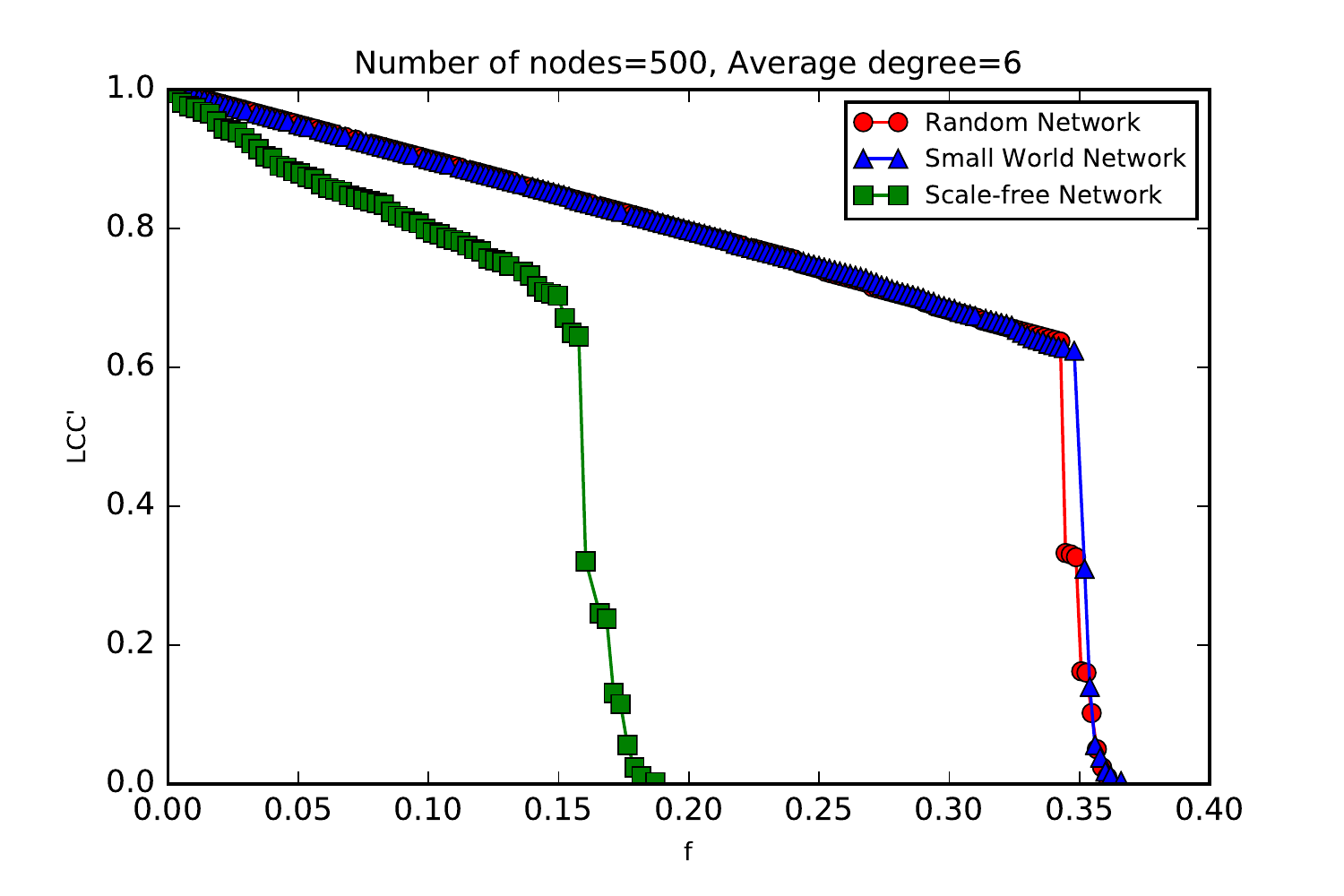}
		\label{fig: Median}}
	\hfil
	\subfloat[RC Attack]{\includegraphics[width=2in]{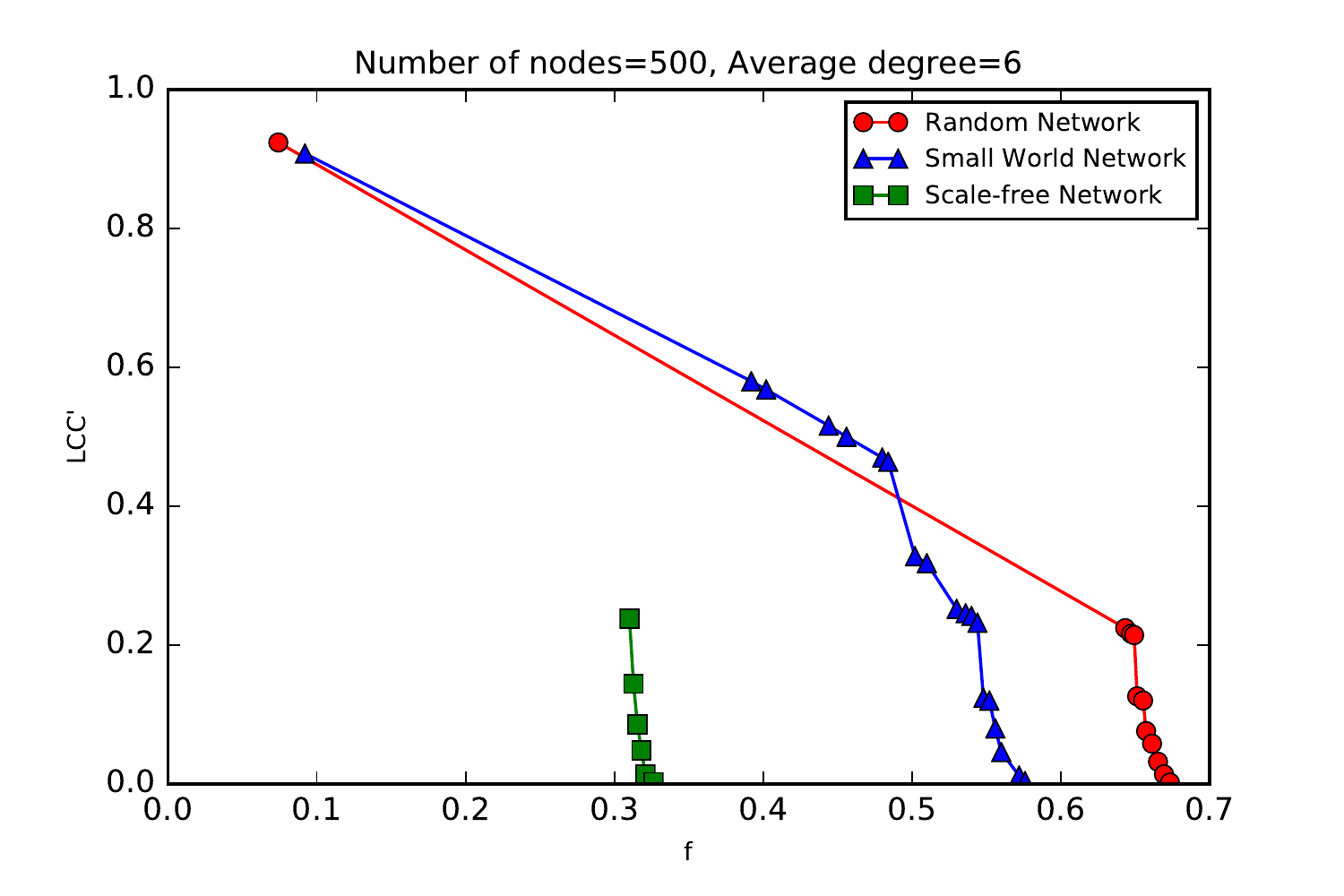}
		\label{fig: Center}}
\\
	\subfloat[RB Attack]{\includegraphics[width=2in]{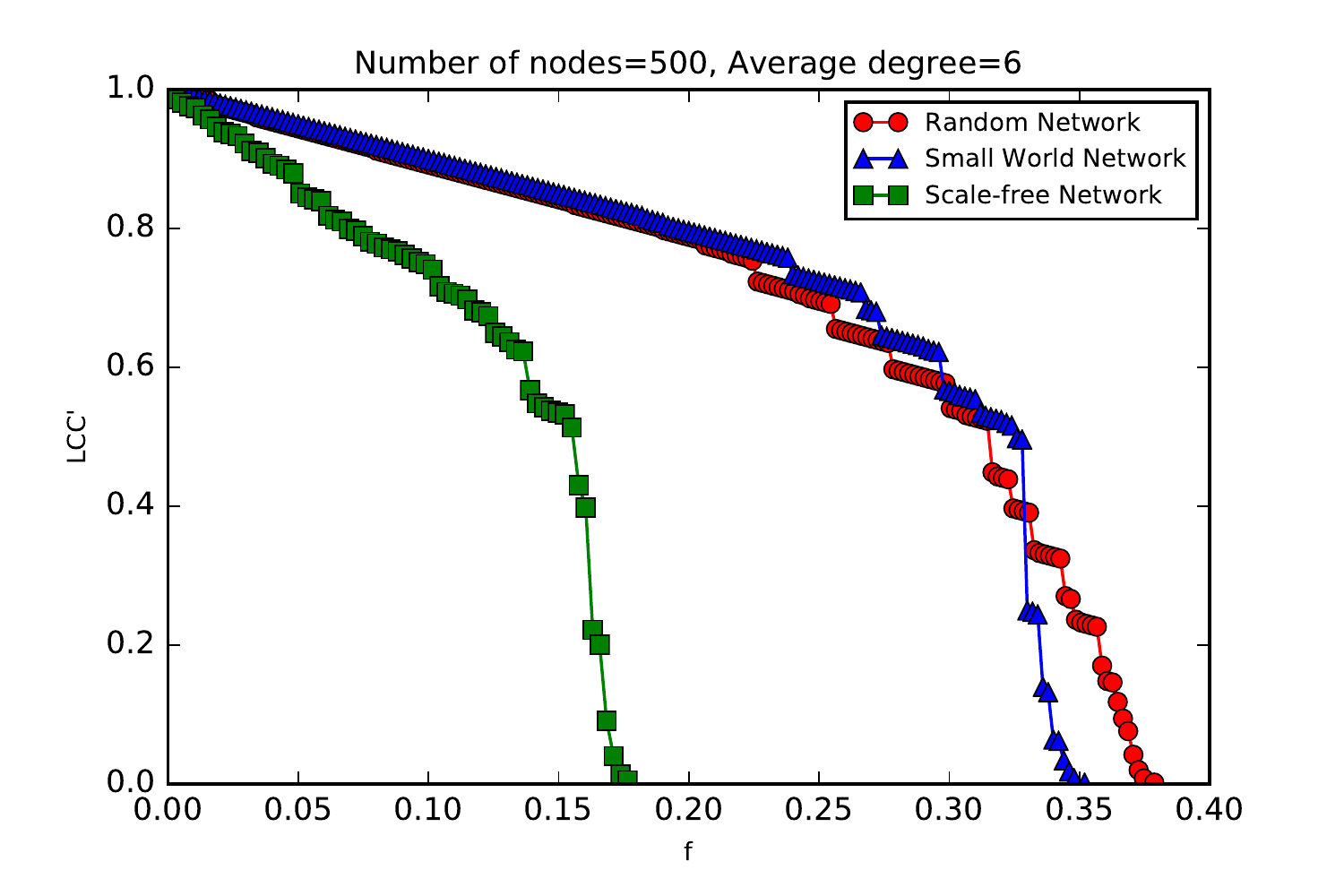}
		\label{fig: BC}}
	\hfil
	\subfloat[RD Attack]{\includegraphics[width=2in]{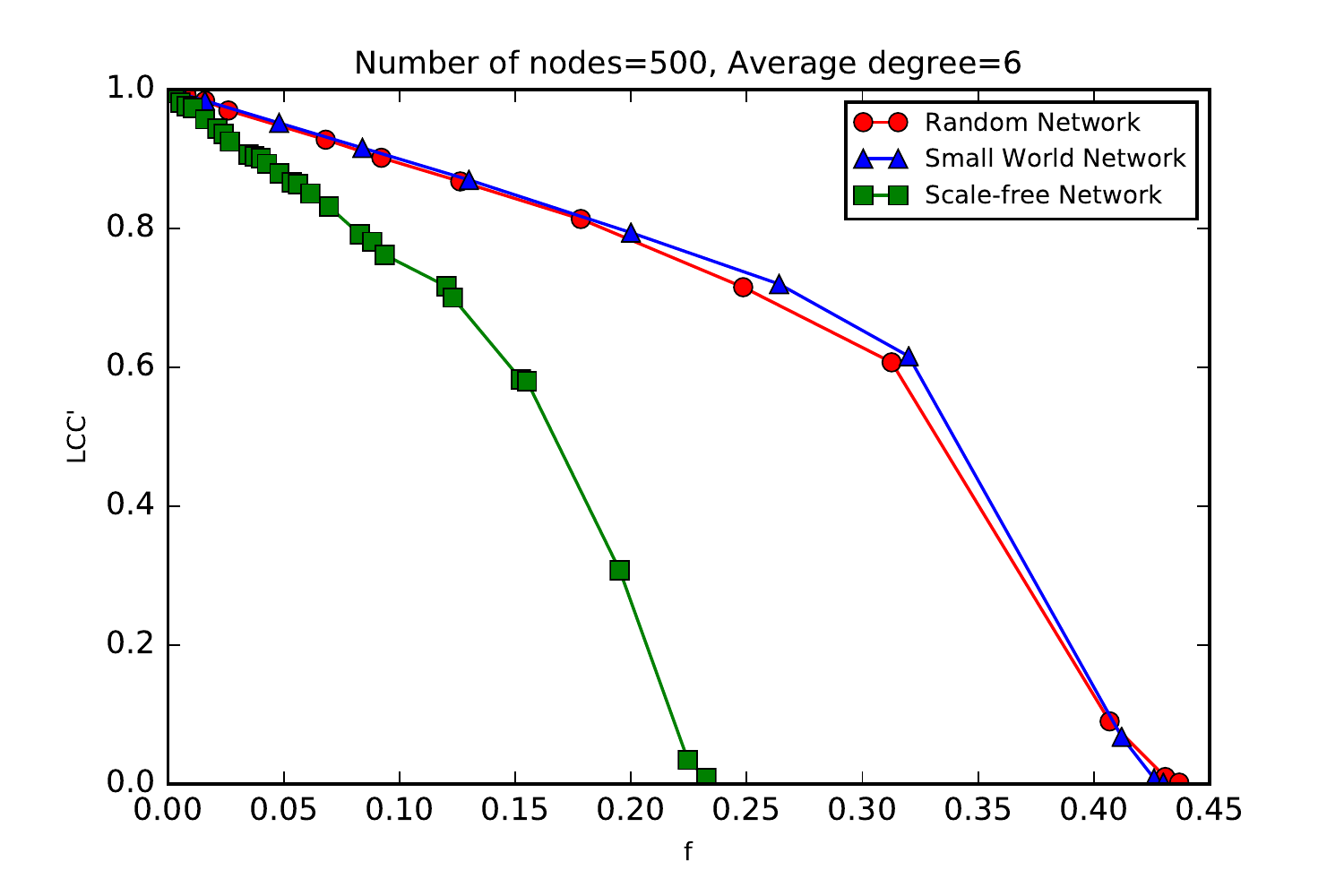}
		\label{fig: DC}}
	\caption{Recalculated attacks in different network models.}
	\label{fig_re}
\end{figure}

\begin{figure}[!t]
	\centering
	\subfloat[Random Network]{\includegraphics[width=2in]{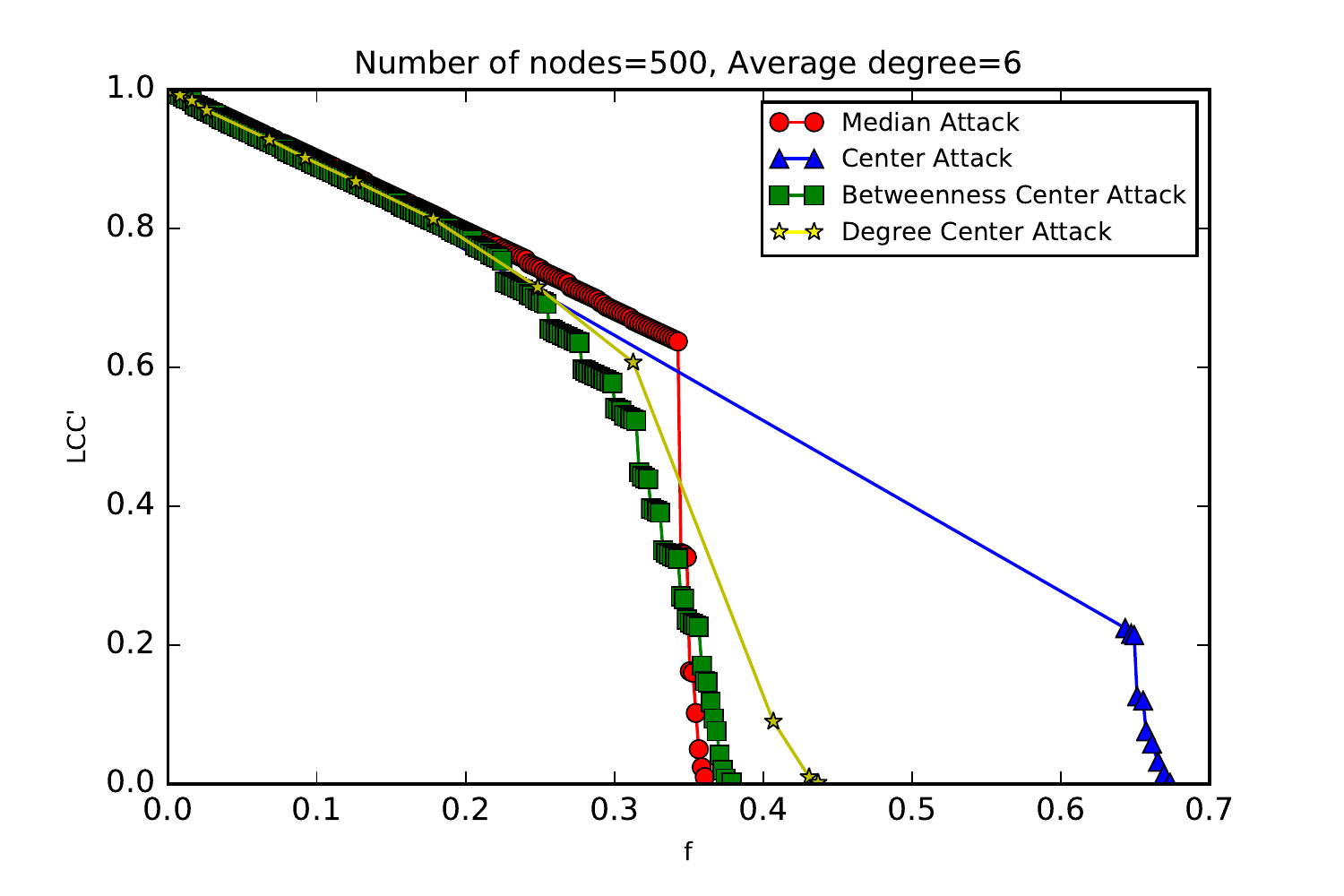}
		\label{fig: Rand}}
	\hfil
	\subfloat[Scale-Free Network]{\includegraphics[width=2in]{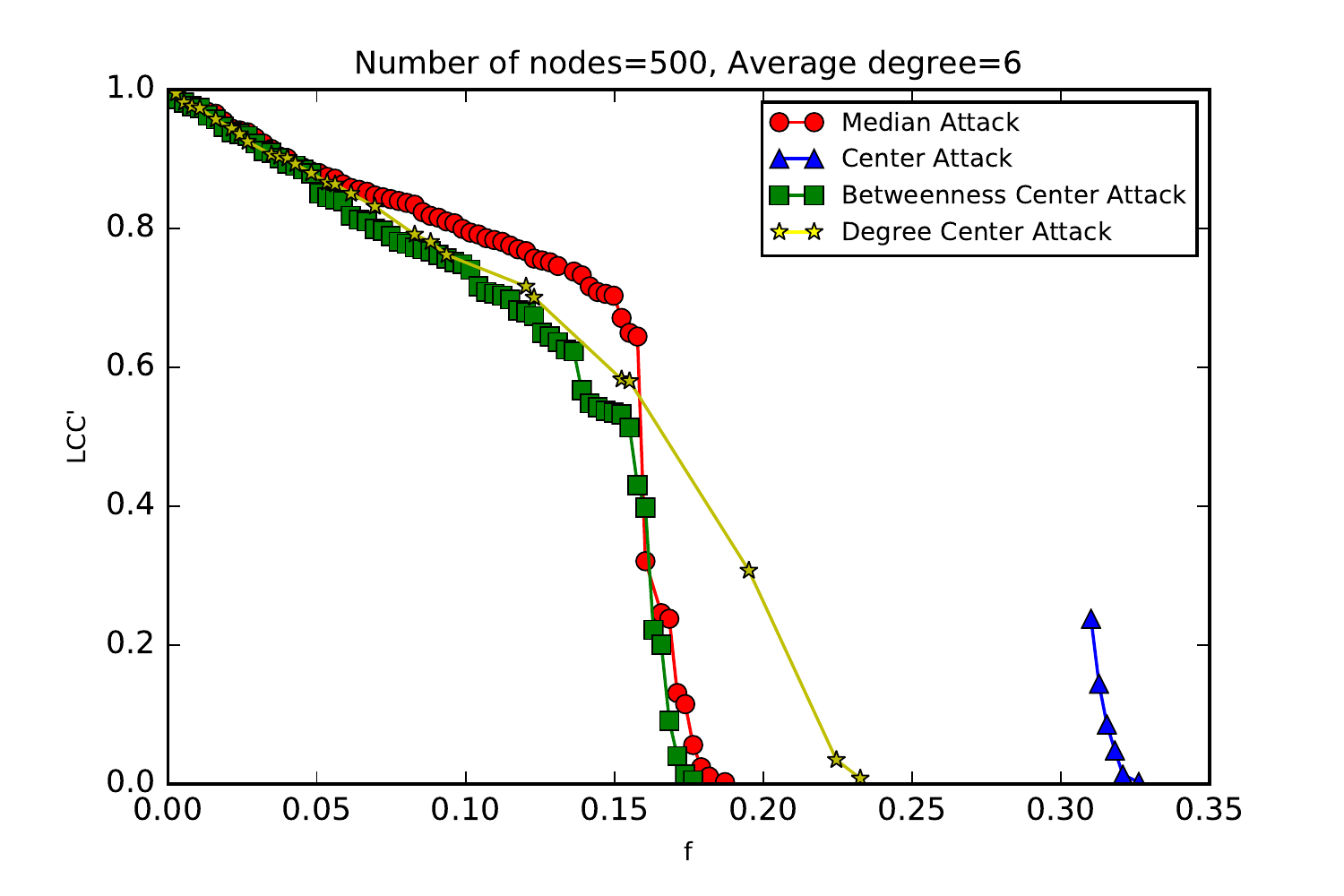}
		\label{fig: SF}}
\\
	\subfloat[Small World Network]{\includegraphics[width=2in]{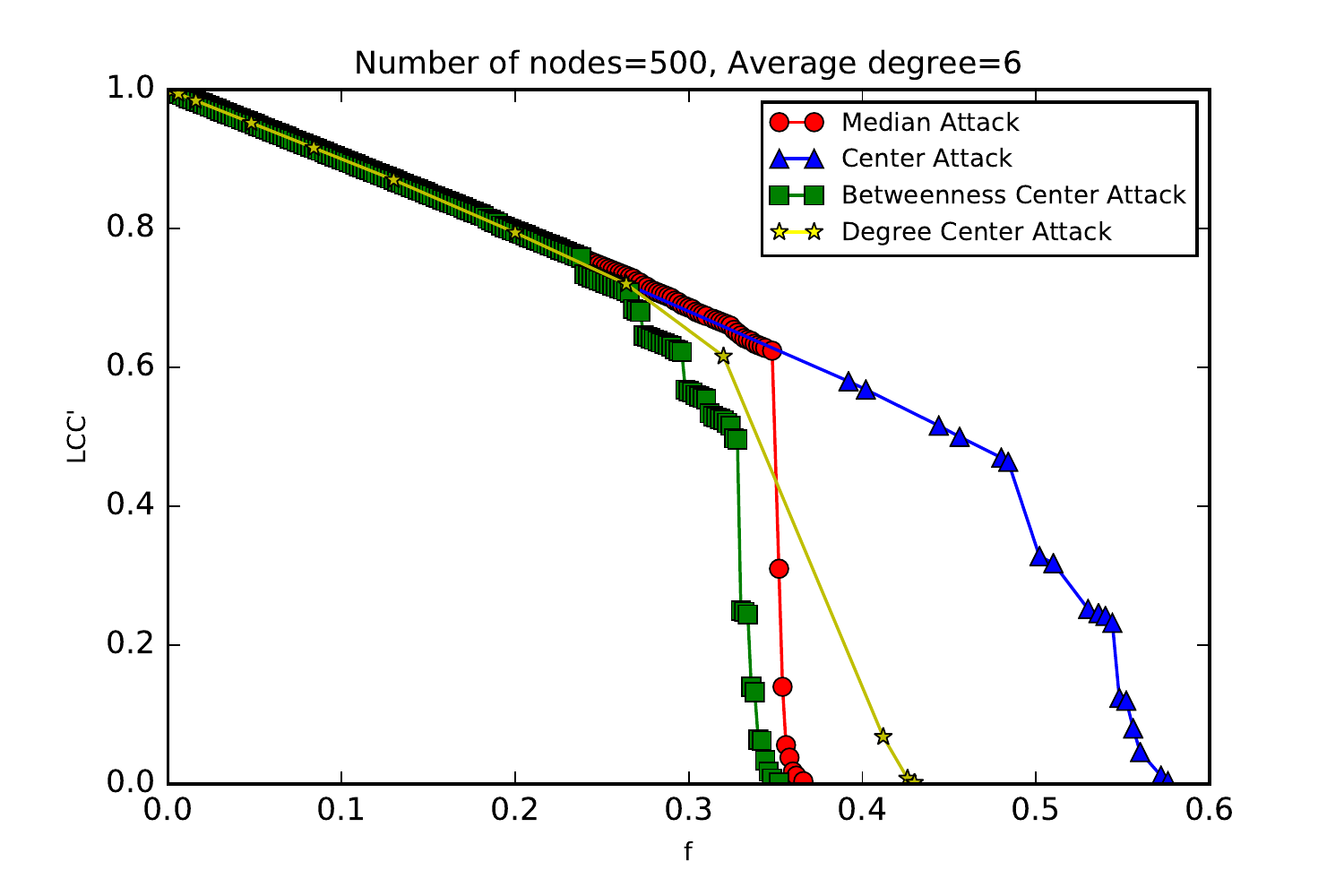}
		\label{fig: SW}}
	\caption{Vulnerability of network models to recalculated attacks.}
	\label{fig_reModel}
\end{figure}

Experiments of attacks based on recalculated information were also carried out as per the strategy described in Algorithm~\ref{algo:cattack}. See figure~\ref{fig_re}. 
The comparison of vulnerability of each network model to different attacks is shown in figure~\ref{fig_reModel}. 

\begin{description}
	\item[Damage measurement] 
	Damage caused by attacks based on both initial and recalculated betweenness center, degree center and median attacks are all the most comparable. Small world and random models were found to be equally robust in these attacks while scale-free networks were easily destructed. Also, in these attacks, networks were completely destructed at very smaller values of $f$ ($\le 0.6$ in initial attacks and $<0.45$ in recalculated attacks) (See figures~\ref{fig_init} and ~\ref{fig_re}). But the response of networks towards $C$ attack was very much different. The networks could withstand the $IC$ attacks up to $f=0.9$ and $RC$ attacks up to $f=0.7$. In this case, also scale-free networks were more vulnerable. Small world networks were completely destructed at comparatively larger $f$ ($\approx 0.9$ for initial attacks and $\approx 0.6$ for recalculated attacks). See figures~\ref{fig: IC} and ~\ref{fig: Center}. However, the tolerance of random models was found to be much better in the case of recalculated attacks ($f \approx 0.7$). (See figure~\ref{fig: Center})
\end{description}
\begin{description}
	\item[Efficiency of attacks] 
	Among the initial center attacks, $IB$ attacks are more efficient in small world and random networks. See figures~\ref{fig: R} and ~\ref{fig: SWi}. In the case of recalculated attacks, all three models are more vulnerable to $RB$ and $RM$ attacks in an almost equal fashion. See figures~\ref{fig: Rand},~\ref{fig: SW},~\ref{fig: SF}. Among them scale-free networks are the weakest ($f< 0.30$ in initial attacks and $f < 0.20$ in recalculated attacks). Comparatively, the $C$ attack is less efficient. All recalculated attacks are obviously more efficient than initial attacks.
\end{description}
\begin{description}
	\item[Change in LCC'] 
	For a given value of $f$, $LCC'$ is very small in scale-free networks compared to small world and random networks (See figures~\ref{fig: Rand},~\ref{fig: SW},~\ref{fig: SF}). This shows that the scale-free networks are less robust than other two models.
\end{description}
\begin{description}
	\item[Network destruction] 
    $IM$, $IB$, $RM$ and $RB$ attacks follow the almost same rate of network destruction. But in $RM$ attack there is a quick and drastic destruction (reduction in $LCC'$) after a threshold value whereas the other three attacks follow a gradual destruction mode (See figures~\ref{fig_init} and ~\ref{fig_re}). 
\end{description}
\textbf{Summary:}  The general observation was that attacks based on median and betweenness are highly hazardous. The efficiency of attacks based on recalculated centers is more than the efficiency of those based on initial network centers. In synthetic networks, the efficiency of attacks was found to be ordered as $IB \ge IM \ge ID \ge IC$ in the case of initial attacks and  $RM \ge RB \ge RD \ge RC$ in the case of recalculated attacks. The most efficient attack among them is $RM$. Among the three network models considered here, the scale-free networks were found to be highly vulnerable towards center-based attacks. This is because there exist a larger number of important vertices in scale-free networks as observed by Albert et.al. in~\cite{AJB00}. They observed that scale-free networks are more sensitive towards node-removal attacks than random or small-world models. Our finding also conforms to this statement.

\section{Vulnerability analysis of real-world networks}\label{sec: Real}

\subsection{Network Science Collaboration Network}
\emph{NetScience network} is a coauthorship network; an undirected network with $1589$ nodes and $2742$ edges. The dataset was downloaded from \href{http://www-personal.umich.edu/~mejn/netdata/}{Newman's page}~\cite{New06}. Nodes in the network represent scientists working on network science. Two nodes are connected if the corresponding scientists have co-authored a publication.

\begin{figure}[!t]
	\centering
	\subfloat[]{\includegraphics[width=2in]{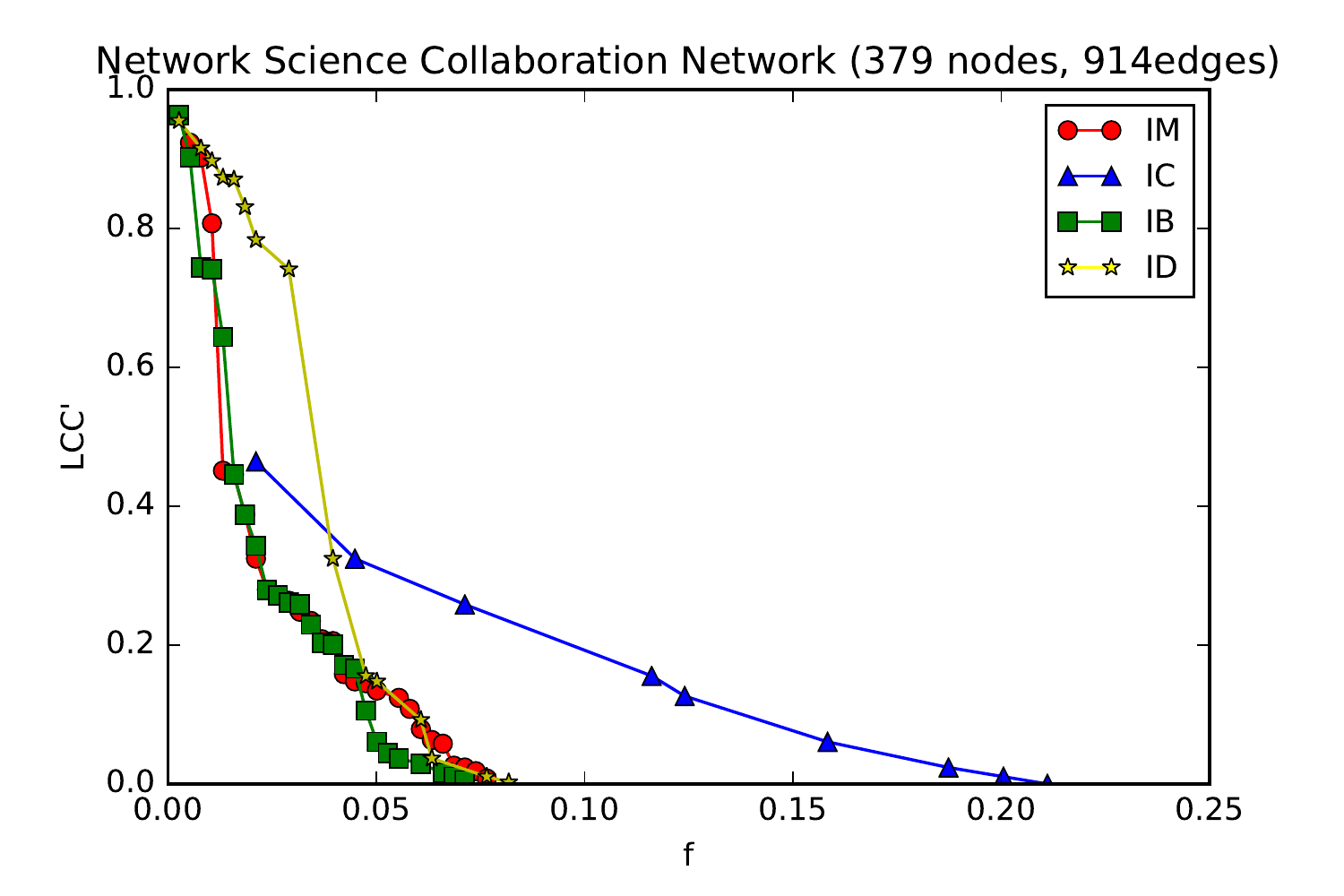}
		\label{fig: INS}}
	\hfil
	\subfloat[]{\includegraphics[width=2in]{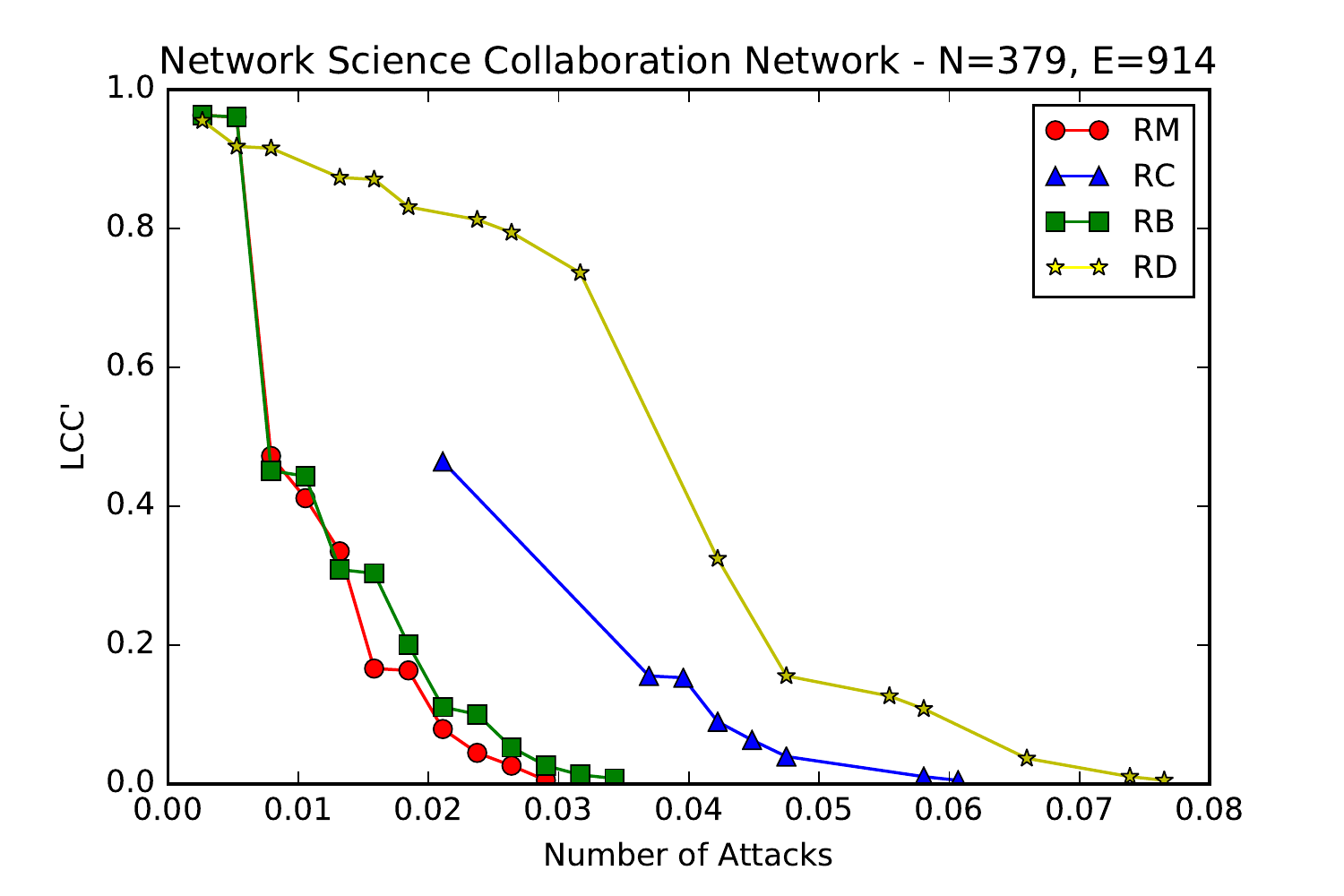}
		\label{fig: RNS}}
	\caption{Initial and recalculated attacks in \emph{NetScience network}.}
	\label{fig: net}
\end{figure}

\emph{Cytoscape 3.5.0}~\cite{SMO03} was used for visualization of the network. Network statistics were obtained using the \emph{Network Analyzer}~\cite{ARS15} of \emph{Cytoscape}. After generating the network we extracted the giant component from it as we focus on the connected network only. Usually, the second largest component in a real complex network is very smaller than the largest (giant) component. This is a characteristic feature of a collaboration network as research will work only if the research community is highly connected~\cite{New01}. In our experiment we consider the giant component of \emph{NetScience network} which consisted of $379$ nodes and $914$ edges. A summary of statistics of the giant component of \emph{NetScience network} is shown in Appendix: Table~\ref{tab:NS}. We analysed the change in $LCC'$ against $f$ in center-based attacks on the giant component of \emph{NetScience network}. See figure~\ref{fig: net}. 
The findings are summarised below:
\begin{itemize}
	\item Initial center attacks like $IM$, $IB$ and $ID$ destructs \emph{NetScience network} in almost same fashion. $IC$ attacks are least efficient among initial value attacks.
	\item The network is highly sensitive to $RM$ and $RB$ attacks ($f<0.04$) and highly robust towards $IC$ attack.
	\item Also, the network is completely destroyed at very lower values of $f$. Even the least efficient attack ($IC$) destroys the network by removing a mere $20$ percent of nodes.
\end{itemize}

\subsection{Les Miserables Coappearance Network}
\begin{figure}[!t]
	\centering
	\subfloat[]{\includegraphics[width=2in]{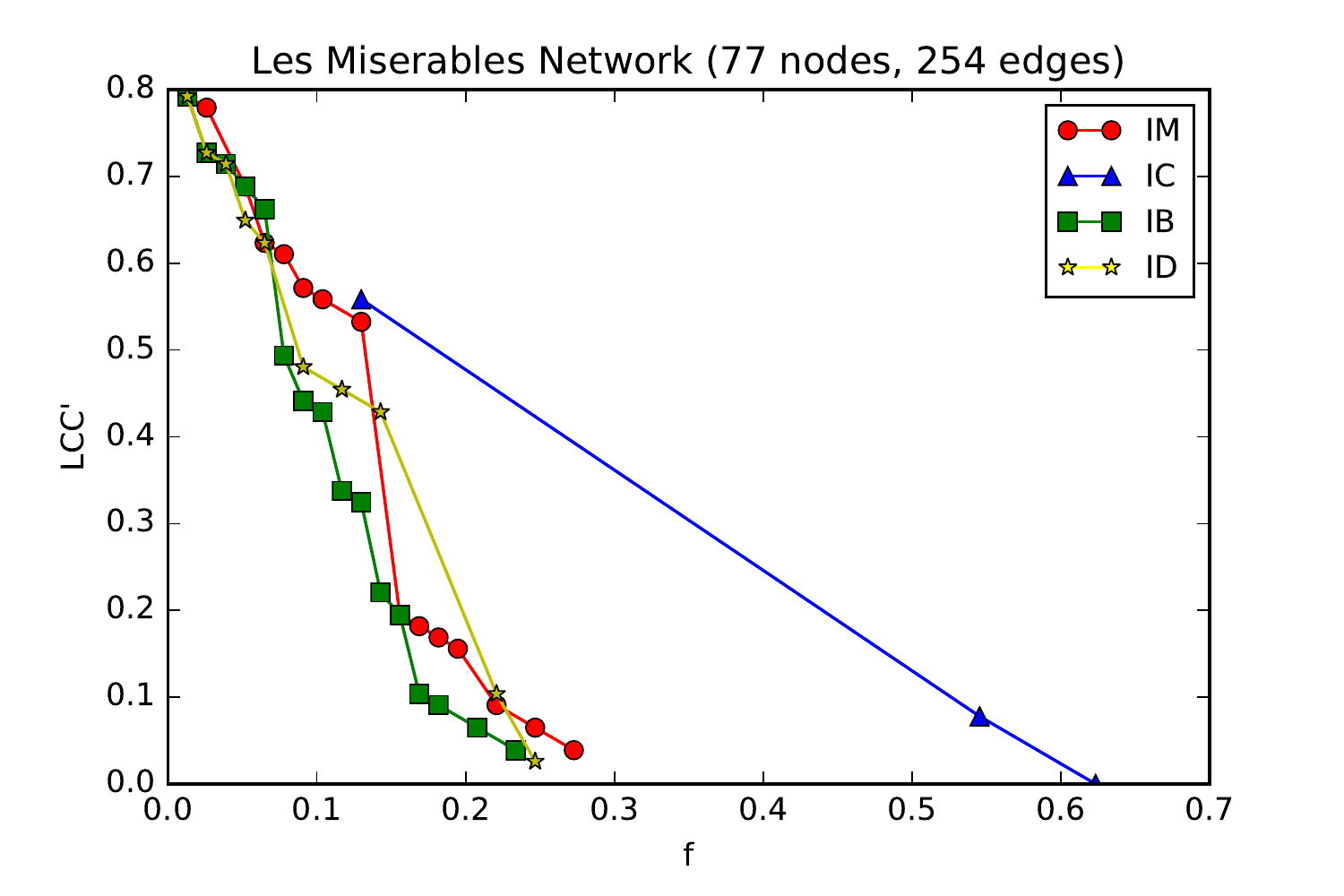}
		\label{fig: ILM}}
	\hfil
	\subfloat[]{\includegraphics[width=2in]{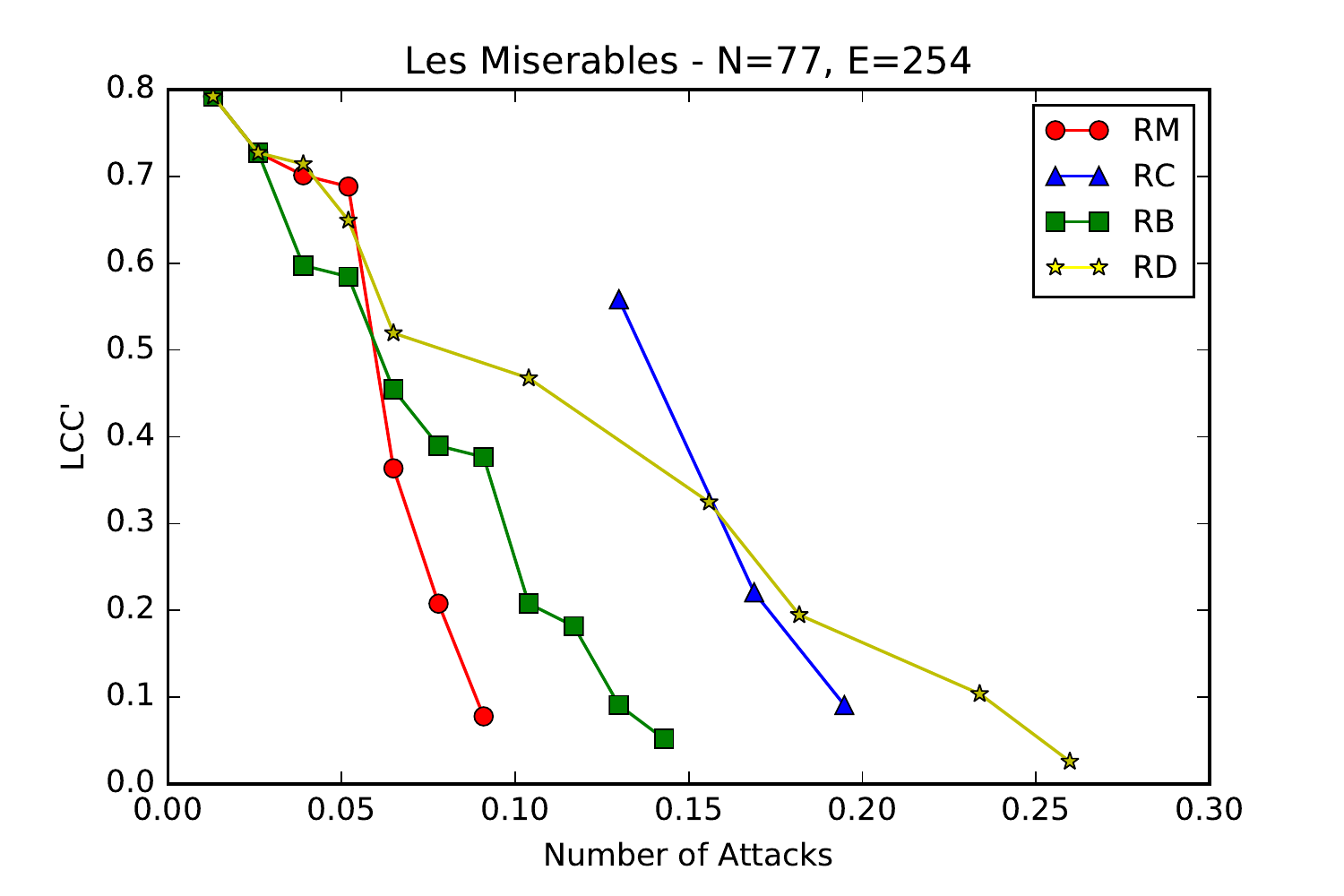}
		\label{fig: RLM}}
	\caption{Initial and recalculated attacks in \emph{LesMis\'{e}rables network}.}
	\label{fig: les}
\end{figure}

This co-appearance network dataset was compiled by Knuth~\cite{Knu93} to represent the co-occurrences of characters in Victor Hugo's famous novel \emph{Les Mis\'{e}rables}. In this undirected network, each node represents a character of the novel and each edge represents the co-occurrence of two characters in the same chapter. For a summary of statistics of \emph{LesMis\'{e}rables network} see Appendix: Table~\ref{tab:LM}. The vulnerability of \emph{LesMis\'{e}rables network} was analysed by running Algorithm~\ref{algo:cattack} and calculating $LCC'$ in each iteration. The plot $LCC'$ vs. $f$ is shown in figure~\ref{fig: les}. 
The findings are summarized below:
\begin{itemize}
	\item \emph{LesMis\'{e}rables network} is most sensitive to $RM$ attack. Complete network is destroyed by removing only $0.1$ fraction of nodes.
	\item Least efficient attack on the network is by $IC$ ($f$ slighlty greater than $0.6$)
	\item Among recalculated attacks, the network shows more robustness towards $RD$ attack.
\end{itemize}

\subsection{US Popular Airport Network}
\begin{figure}[!t]
	\centering
	\subfloat[]{\includegraphics[width=2in]{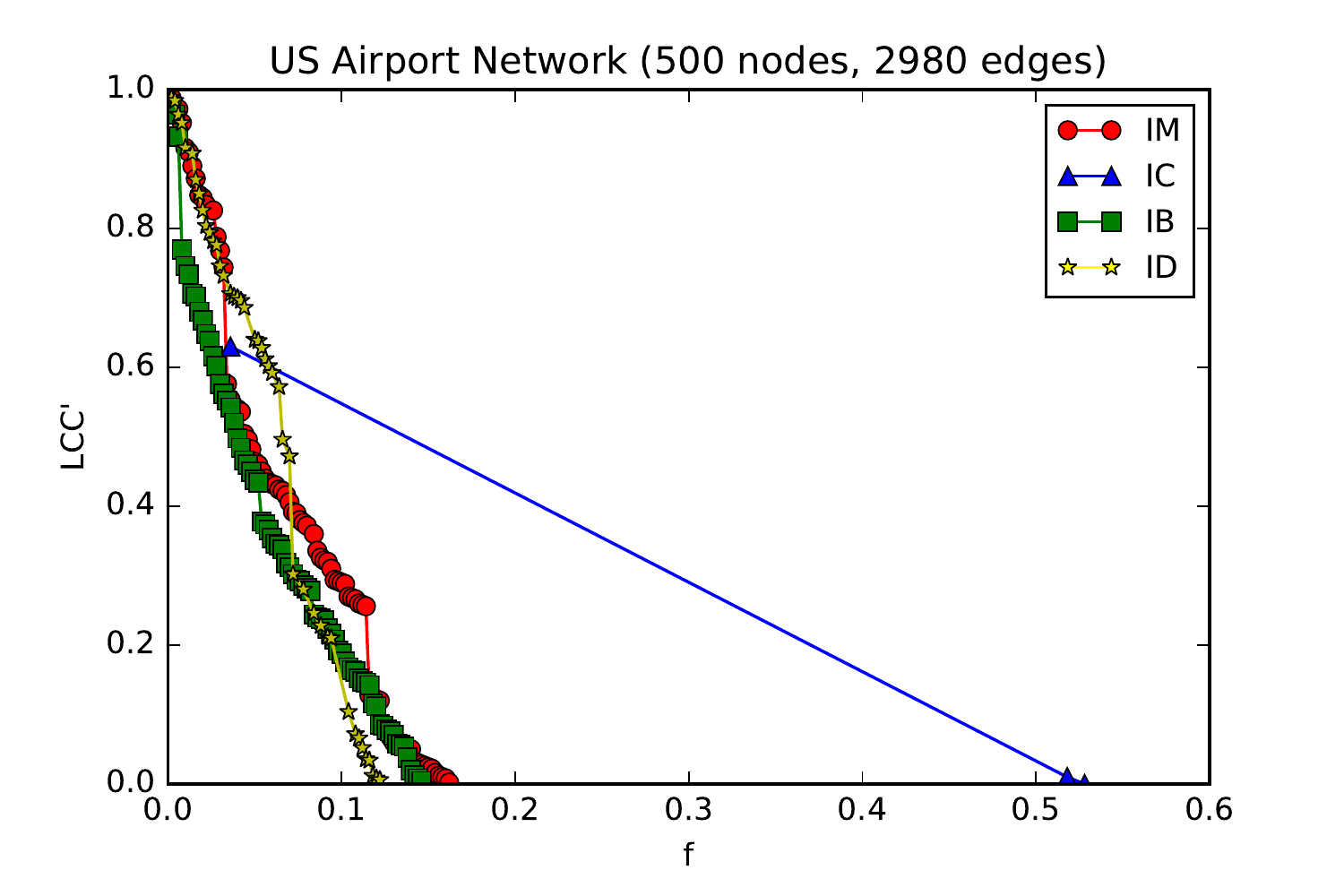}
		\label{fig: IUS}}
	\hfil
	\subfloat[]{\includegraphics[width=2in]{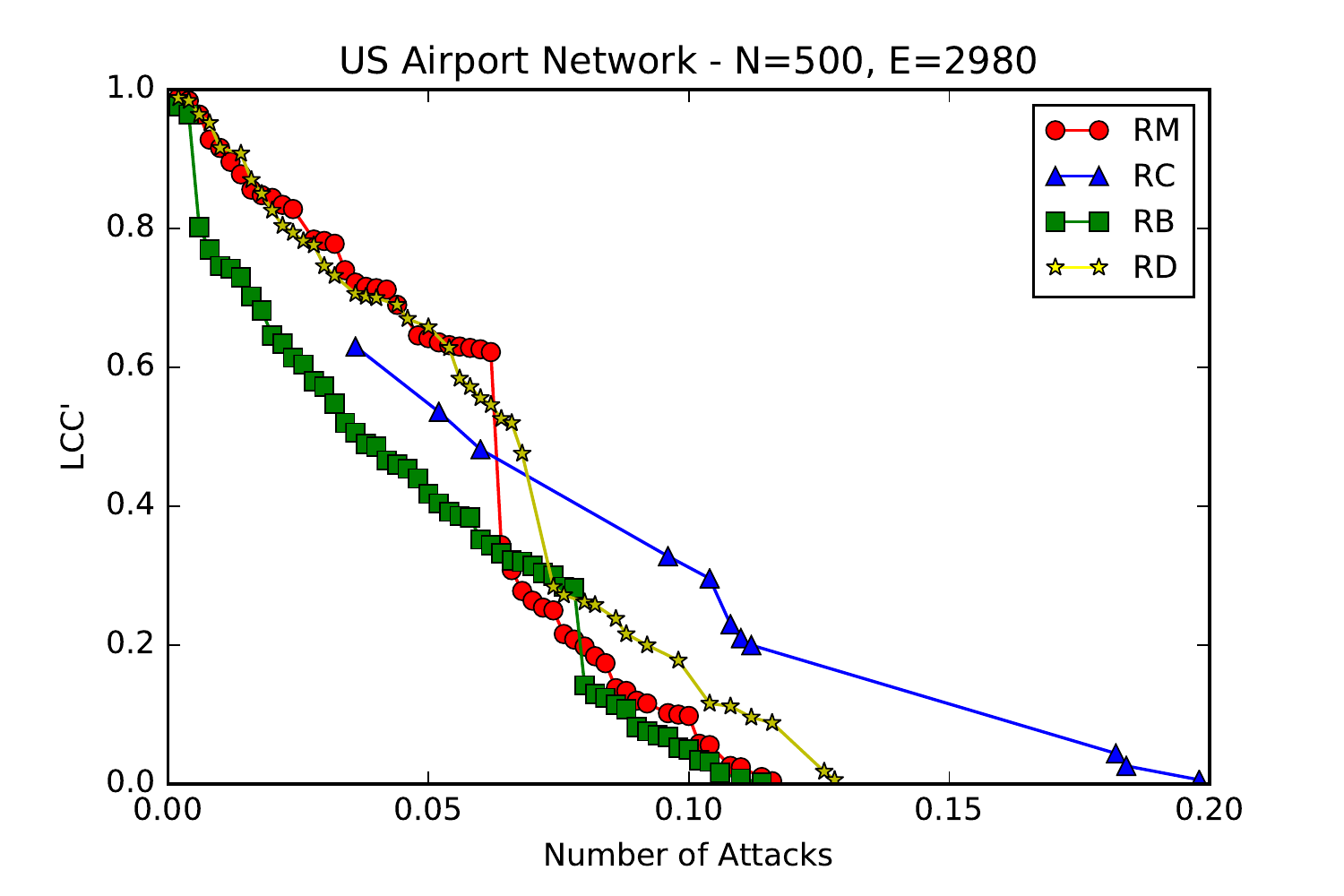}
		\label{fig: RUS}}
	\caption{Initial and recalculated attacks in \emph{US Airport network}.}
	\label{fig: US}
\end{figure}

This is an undirected network of 500 most popular US airports. The dataset was compiled and used by Colizza et. al. in 2007~\cite{CPV07}. The nodes represent airports and edges represent the air traffic between them. Hereafter the network is called as \emph{US Airport network}. See Appendix: Table~\ref{tab:US} for a summary of statistics of \emph{US Airport network}. After running Algorithm~\ref{algo:cattack} on \emph{US Airport network} we analysed the change in $LCC'$ against $f$. See figure~\ref{fig: US}.
The observations from the experiment are as follows:
\begin{itemize}
	\item \emph{US Airport network} is more sensitive to $RM$ and $RB$ attacks.
	\item The network shows more robustness to $IC$ attack.
\end{itemize}

\subsection{Yeast Protein Interaction Network}
\begin{figure}[!t]
	\centering
	\subfloat[]{\includegraphics[width=2in]{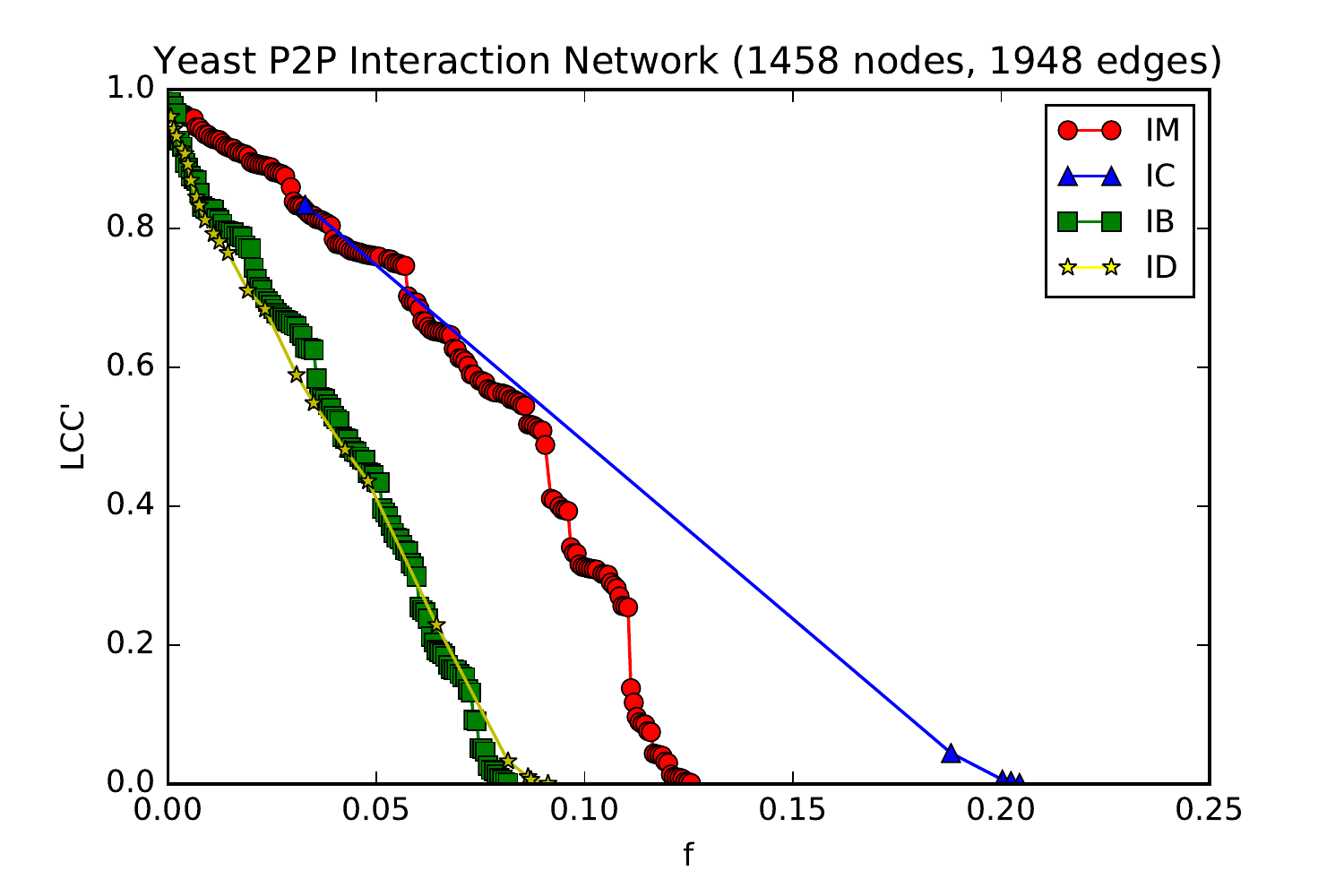}
		\label{fig: IYS}}
	\hfil
	\subfloat[]{\includegraphics[width=2in]{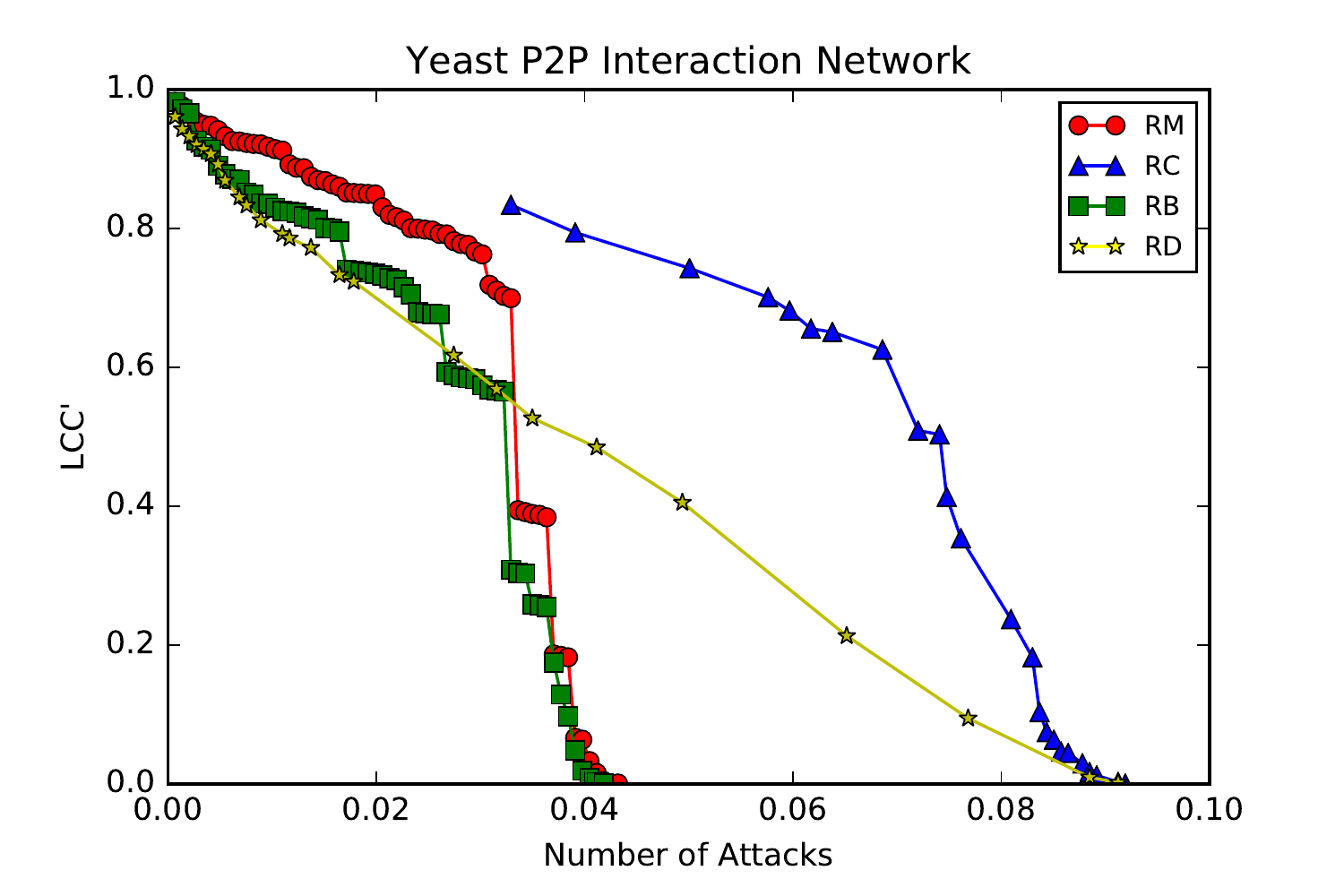}
		\label{fig: RYS}}
	\caption{Initial and recalculated attacks in \emph{Yeast Protein Interaction network}.}
	\label{fig: yst}
\end{figure}

The biological network considered in this study is the protein interaction network for the yeast Saccharomyces cerevisiae. This data set was used by Jeong et. al. in their paper~\cite{JMB01}. The nodes represent proteins and edges represent protein-protein interaction. The actual network consists of 2114 nodes and 2277 edges. But here we extracted and used only the giant component (of 1458 nodes) for our study. See Appendix: Table~\ref{tab:YS} for a summary of statistics of the giant component of \emph{Yeast Protein Interaction network}. See figure~\ref{fig: yst} for the plot $LCC'$ vs. $f$ during center-based attacks on \emph{Yeast Protein Interaction network}.
The findings are as follows:
\begin{itemize}
	\item Attacks on \emph{Yeast Protein Interaction network} are comparable to those on \emph{NetScience network}. Like the latter network, it is more sensitive to $RM$ and $RB$ attacks with network destroyed at very small values of $f$ ($\approx 0.04$).
	\item This network also shows more robustness to $IC$ attack like all other networks.
\end{itemize}
\textbf{Summary: }The efficiency of attacks in real-world networks was found to be as $IB \approx IM \approx ID > IC$ in the case of initial center attacks and $RM > RB > RC \textnormal{ or } RD$ in the case of recalculated center attacks. Among the recalculated attacks, the collaboration network and literature network shows more robustness towards $RD$ attack while air-transport network and biological network are highly tolerant towards $RC$ attack. Among the different real-world networks considered here the robustness of collaboration network was found to be highly weak. 

\section{Conclusion}
We proposed center-based attack strategies and analysed the tolerance of different synthetic and real-world networks towards these attacks. Network attacks based on betweenness center and degree center were studied in detail in papers like ~\cite{CLM04, GHR11, HHL16, HKY02, IKS13, MBK11}. However, comparison of network attacks based on different centers is a naive approach. The attacks based on center and median are seldom found in the literature. An important milestone of this study is the introduction of median-based attack. Attacks based on median are found to be highly efficient among other center-based attacks. Nodes with higher median value are those with minimum total distance in the network. Therefore identifying and removing such nodes increase the value of geodesics in the network and in turn induce heavy damages to the network structure.

\emph{Future Scope} An interesting follow-up to this work is to study network survivability in center-based attacks. Robustness of a network can be analysed more accurately by accounting for weights in the network. This also opens up a new direction for further follow-up investigation.
\section*{Funding}
This work was supported by the National Post Doctoral Fellowship (N-PDF) No. PDF/2016/002872 from Science and Engineering Research Board (SERB), Department of Science and Technology (DST), Government of India.
\section*{Acknowledgements}
The authors would like to thank Prof. M. Bellingeri for providing review comments on a previous draft of the paper.
\appendix
\section*{Appendix}
\begin{table}[H]
	\caption{Statistical parameters: Giant component of \emph{NetScience network}}
	\label{tab:NS}	
	\centering	
	\begin{tabular}{ | p{10cm} | p{1cm}  |}
		\hline
		\multicolumn{1}{|c|}{\textbf{Network Parameter}} & \multicolumn{1}{c|}{\textbf{Value}}\\
		\hline
		\hline
		Number of authors with collaborations (Node Count)& \multicolumn{1}{c|}{$379$} \\
		\hline
		Number of connected components (ncc)& \multicolumn{1}{c|}{$1$} \\
		\hline
		Shortest distance between farthest authors (Diameter)& \multicolumn{1}{c|}{$17$} \\
		\hline
		Shortest distance between closest authors (Radius)& \multicolumn{1}{c|}{$9$} \\
		\hline
		Number of connected author-pairs (Shortest paths)& \multicolumn{1}{c|}{$143262$} \\
		\hline
		Expected distance between two connected authors (Characteristic path length)& \multicolumn{1}{c|}{$6.042$} \\
		\hline
		Average Number of co-authors (Average degree)& \multicolumn{1}{c|}{$4.823$} \\
		\hline
		Average chance of collaboration between co-authors (Clustering coefficient)& \multicolumn{1}{c|}{$74.12\%$} \\
		\hline
	\end{tabular}
\end{table}
\bigskip
\begin{table}[H]
	\caption{Statistical parameters: \emph{LesMis\'{e}rables network}}
	\label{tab:LM}
	\centering
	\begin{tabular}{ | p{10cm} | p{1cm}  |}
		\hline
		\multicolumn{1}{|c|}{\textbf{Network Parameter}} & \multicolumn{1}{c|}{\textbf{Value}}\\
		\hline
		\hline
		Number of characters in the novel (Node count)& \multicolumn{1}{c|}{$77$} \\
		\hline
		Number of connected components (ncc)& \multicolumn{1}{c|}{$1$} \\
		\hline
		Shortest distance between farthest characters (Diameter)& \multicolumn{1}{c|}{$5$} \\
		\hline
		Shortest distance between closest characters (Radius)& \multicolumn{1}{c|}{$3$} \\
		\hline
		Number of connected character-pairs (Shortest paths)& \multicolumn{1}{c|}{$5852$} \\
		\hline
		Expected distance between two connected characters (Characteristic path length)& \multicolumn{1}{c|}{$2.641$} \\
		\hline
		Average number of co-appearing characters (Average degree)& \multicolumn{1}{c|}{$6.597$} \\
		\hline
		Average chance of co-appearance between co-appearing characters (Clustering coefficient)& \multicolumn{1}{c|}{$57.313\%$} \\
		\hline
	\end{tabular}
\end{table}
\bigskip
\begin{table}[H]
	\caption{Statistical parameters: \emph{US Airport network}}
	\label{tab:US}
	\centering
	\begin{tabular}{ | p{10cm} | p{1cm}  |}
		\hline
		\multicolumn{1}{|c|}{\textbf{Network Parameter}} & \multicolumn{1}{c|}{\textbf{Value}}\\
		\hline
		\hline
		Number of airports (Node count)& \multicolumn{1}{c|}{$500$} \\
		\hline
		Number of connected components (ncc)& \multicolumn{1}{c|}{$1$} \\
		\hline
		Shortest distance between farthest airports (Diameter)& \multicolumn{1}{c|}{$7$} \\
		\hline
		Shortest distance between closest airports (Radius)& \multicolumn{1}{c|}{$4$} \\
		\hline
		Number of connected airport-pairs (Shortest paths)& \multicolumn{1}{c|}{$250500$} \\
		\hline
		Expected distance between two connected airports (Characteristic path length)& \multicolumn{1}{c|}{$2.999$} \\
		\hline
		Average number of directly connected airports(Average degree)& \multicolumn{1}{c|}{$11.896$} \\
		\hline
		Average chance of direct connection between airports connected to one airport (Clustering coefficient)& \multicolumn{1}{c|}{$61.357\%$} \\
		\hline
	\end{tabular}
\end{table}
\bigskip
\begin{table}[H]
	\caption{Statistical parameters: Giant component of \emph{Yeast Protein Interaction network}}
	\label{tab:YS}
	\centering
	\begin{tabular}{ | p{10cm} | p{1cm}  |}
		\hline
		\multicolumn{1}{|c|}{\textbf{Network Parameter}} & \multicolumn{1}{c|}{\textbf{Value}}\\
		\hline
		\hline
		Number of proteins (Node count)& \multicolumn{1}{c|}{$1458$} \\
		\hline
		Number of connected components (ncc)& \multicolumn{1}{c|}{$1$} \\
		\hline
		Shortest distance between farthest proteins (Diameter)& \multicolumn{1}{c|}{$19$} \\
		\hline
		Shortest distance between closest proteins (Radius)& \multicolumn{1}{c|}{$11$} \\
		\hline
		Number of connected protein-pairs (Shortest paths)& \multicolumn{1}{c|}{$2124306$} \\
		\hline
		Expected distance between two interacting proteins (Characteristic path length)& \multicolumn{1}{c|}{$6.812$} \\
		\hline
		Average number of directly interacting proteins (Average degree)& \multicolumn{1}{c|}{$2.672$} \\
		\hline
		Average chance of interaction between proteins interacting with same protein (Clustering coefficient)& \multicolumn{1}{c|}{$70.83\%$} \\
		\hline
	\end{tabular}
\end{table}

\end{document}